\documentstyle[twocolumn]{mn}
\input{psfig.sty}
\title[Giant radio sources]
{Giant radio sources}

\author[C.H. Ishwara-Chandra and  D.J. Saikia]
{C.H. Ishwara-Chandra$^{1\,,2}$\thanks{E-mail: ishwar@ncra.tifr.res.in} and 
D.J. Saikia$^{1}$\thanks{E-mail: djs@ncra.tifr.res.in}  \\
$^1$ National Center for Radio Astrophysics, TIFR, Post Bag 3, 
Ganeshkhind, Pune 411 007, India \\
$^2$ Joint Astronomy Programme, Department of Physics, Indian 
Institute of Science, Bangalore, 560 012, India
}
 
\date{}
\begin{document}
\maketitle
\begin{abstract}
We present multi-frequency VLA observations of two giant quasars, 0437$-$244 and 
1025$-$229, from the Molonglo Complete Sample. These sources have well-defined
FRII radio structure, possible one-sided jets, no significant depolarization
between 1365 and 4935 MHz and low rotation measure ($\mid$ RM $\mid$ $<\,$ 20 
rad m$^{-2}$). The giant sources are defined to be those whose overall 
projected size is $\geq$ 1 Mpc.
We have compiled a sample of about 50 known giant radio sources from the literature,
and have compared some of their properties with a complete sample of 3CR
radio sources of smaller sizes to investigate the evolution of giant sources,
and test their consistency with the unified scheme for radio galaxies and
quasars. We find an inverse correlation between the degree of core prominence 
and total radio luminosity, and
show that the giant radio sources have similar core strengths to the smaller sources 
of similar total luminosity. Hence their large sizes are unlikely to be due to stronger
nuclear activity. The degree of collinearity of the giant sources is also similar to
the sample of smaller sources. The luminosity-size diagram shows that the giant 
sources are 
less luminous than our sample of smaller-sized 3CR sources, consistent with evolutionary
scenarios where the giants have evolved from the smaller sources losing energy
as they expand to these large dimensions. For the smaller sources, radiative losses due
to synchrotron radiation is more significant while for the giant sources the 
equipartition magnetic fields are smaller and inverse Compton losses with the
microwave background radiation is the dominant process. The radio properties of the
giant radio galaxies and quasars are consistent with the unified scheme.
\end{abstract}


\section{Introduction}
Giant radio sources (GRSs), defined to be those with a projected linear size
$\geq$1 Mpc (q$_\circ$=0.5 and H$_\circ$=50 km s$^{-1}$ Mpc$^{-1}$),
are the largest single objects in the Universe, and are extremely useful for 
studying a number of astrophysical problems. These range from understanding 
the evolution of radio sources, constraining orientation-dependent
unified schemes to probing the intergalactic medium at different redshifts 
(e.g. Saripalli 1988; Subrahmanyan \& Saripalli 1993, Subrahmanyan, Saripalli \& 
Hunstead 1996;
Mack et al. 1998; Schoenmakers et al. 1998a). There are about 50 known giant sources,
only 5 of which are quasars, the largest being the radio galaxy 3C236 with 
a projected linear size of 5.7 Mpc (Willis, Strom \& Wilson  1974; 
Strom \& Willis 1980; Barthel et al. 1985). In the complete sample of 3CR radio
sources (Laing, Riley \& Longair 1983), about 6 per cent of the radio sources
are giants. The GRSs are usually of low radio
luminosity with values often in the transition region between the FRI and FRII
type sources (Fanaroff \& Riley 1974), and are believed to be advancing outwards
through the low-density (10$^{-5} - 10^{-6}$ cm$^{-3}$) intergalactic medium. 
Their spectral ages have been estimated to be about 10$^{7}$ to 10$^{8}$ yr
(e.g. Mack et al. 1998), but these values are dependent on a number of assumptions
and have to be treated with caution (e.g. Eilek, Melrose \& Walker 1997).

In this paper we report further observations of two giant quasars (GQs), 0437$-$244
and 1025$-$229, which we had earlier observed as part of our study of the depolarization
properties of a well-defined sample of radio sources selected from the Molonglo
Complete sample (Ishwara-Chandra et al. 1998, hereinafter referred to as IC98, and 
references therein). The quasar
0437$-$244 at a redshift of 0.84 is at present the highest redshift giant quasar known.
The paper is organized as follows.  In Section 2, observations of the two giant quasars 
from the Molonglo Sample are presented. 
In Section 3, we list some of the properties of known giant radio galaxies and
quasars, and discuss their consistency with  proposed evolutionary 
scenarios for these objects.
In Section 4, we examine the consistency of the giant sources with the unified schemes 
for radio sources, and possible effects of the environment. 
The conclusions are summarized in Section 5.

\section{Giant quasars from the Molonglo sample}
In this section we present the observations and observational results of
the two giant quasars 0437$-$244 and 1025$-$229 from the Molonglo complete
sample. The GQ 0437$-$244 has an angular size of 128$''$ which corresponds to
a projected linear size of 1.06 Mpc at a redshift of 0.84. This is presently the highest 
redshift GQ; the only other known GQ with a redshift greater than 0.5 is 1127$-$300 
(Bhatnagar,  Gopal-Krishna \& Wisotzki 1998) which is at a redshift of 0.6337. The second GQ 1025$-$229 is 
relatively nearby at a redshift of
0.309, and has an angular size of 198$''$, which corresponds to a linear size of 1.11 Mpc.
The luminosities at 1.4 GHz are 2.88 $\times 10^{27}$ and
3.47 $\times 10^{26}$ W Hz$^{-1}$ for 0437$-$244 and 1025$-$229 respectively.

\subsection{Observations and analyses}
The sources were observed in the L, C, X and U bands with the 
Very Large Array (VLA) and the observing log is summarized in Table 1.
The observations in the L, C and U bands were scaled-array ones, while
those in the X-band were with a higher resolution. The L- and C-band 
total-intensity and linear-polarization images
have been reported earlier by IC98. However, the
spectral index and rotation measure estimates 
based on these images are presented here. All  the data were
calibrated in the standard way using the NRAO {\tt AIPS} package. 
The final images in the L, C and U bands were restored with a 
beam of 4.5$'' \times 3.2''$ along PA = $-70^\circ$ for  0437$-$244, and 
8.0$'' \times 4.5''$ along PA $-50^\circ$ for 1025$-$229.
The restoring beam for the BnA-array observations at the X-band are 
1.89$'' \times 0.58''$ along PA $35^\circ$ for 0437$-$244,
and 0.80$'' \times 0.61''$ along PA $53^\circ$ for 1025$-$229. The images
at the X and U bands have been corrected for primary beam attenuation. The
U-band images have not been self-calibrated because they have low signal to
noise ratio.

\begin{table} \caption{ Observing log }
\begin{tabular}{l l l l l }

Array  & Obs.  &Obs.     & Band-  & Date of obs.  \\
Conf.  & band  &Freq.    & width  &        \\
       &       & MHz     & MHz    &        \\
       &       &         &        &     \\
BnA    & L     &1365   &  50 & 1995 Sep 20   \\
       & L     &1665   &  25 &                 \\
CnB    & C     &4635   &  50 & 1996 Jan 20,31  \\
       & C     &4935   &  50 &                 \\        
DnC    & U     &14965  &  50 &  1997 Oct 3, 12    \\
BnA    & X     &8447   &  25 &  1997 Feb 3  \\
\end{tabular}
\end{table}

\subsection{Observational results}

\subsubsection{0437$-$244}
In Figure 1a we present the total-intensity image at 1365 MHz with the spectral index 
between 1365 and 4935 MHz superimposed in grey. 
The scaled-array 15 GHz image, and the high-resolution 8.4 GHz image 
are presented in Figures 1b and 1c respectively. 

\begin{figure*}
\vbox{
\vspace{-0.075 in}
\hbox{
\hspace{0.25 in}
\psfig{figure=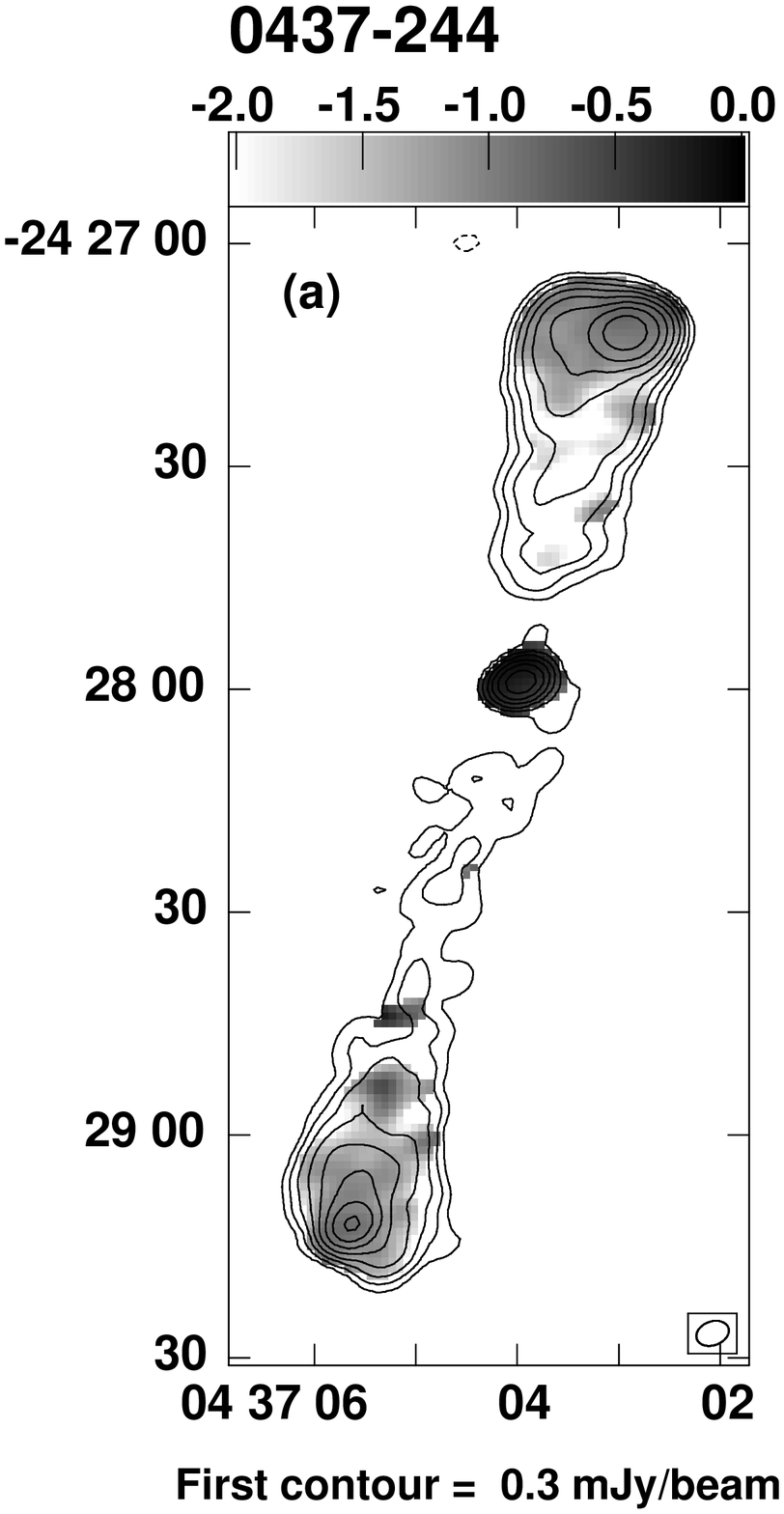,height=4.1in}
\hspace{0.3 in}
\psfig{figure=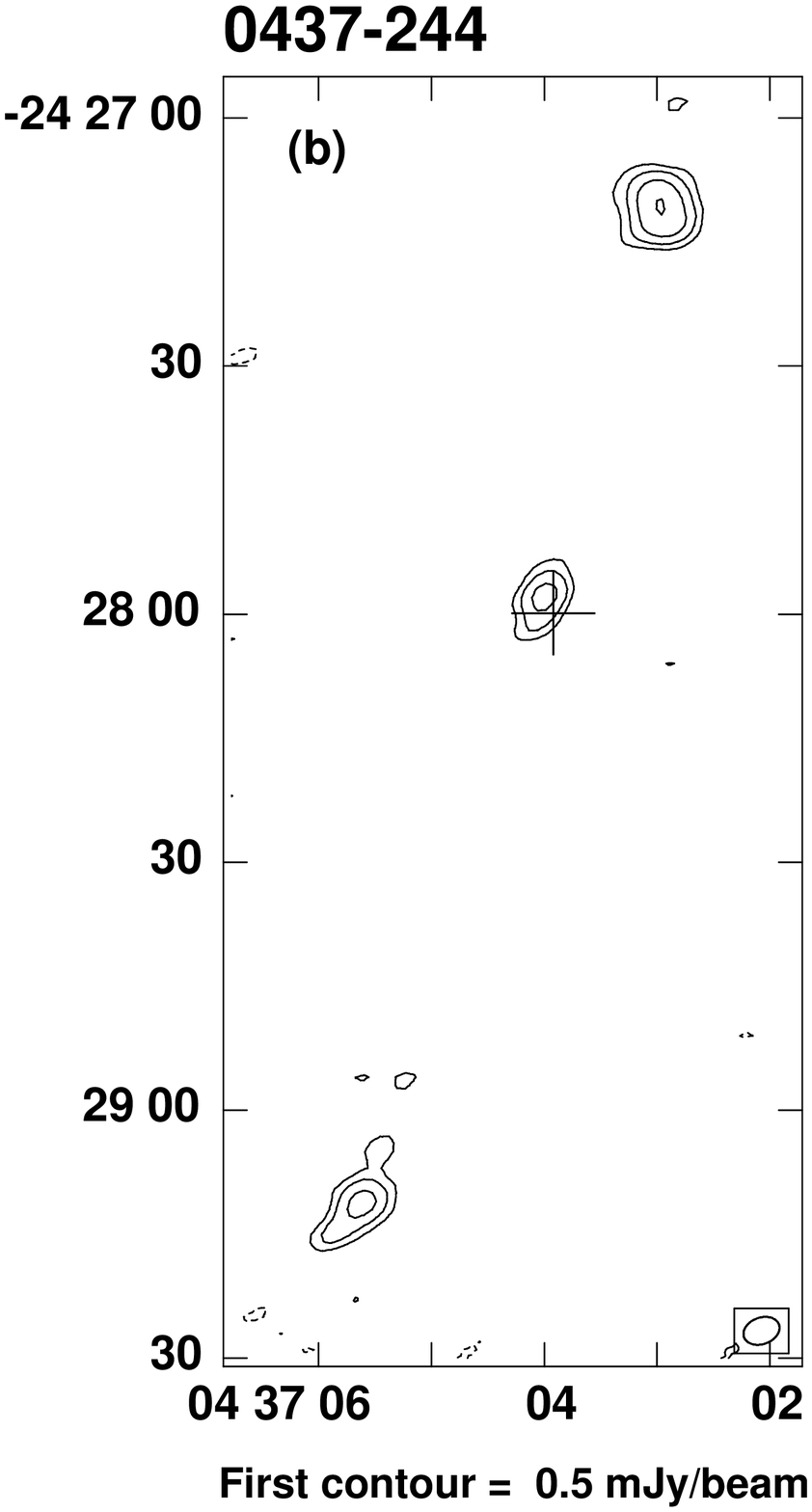,height=4.0in}
\hspace{0.3 in}
\psfig{figure=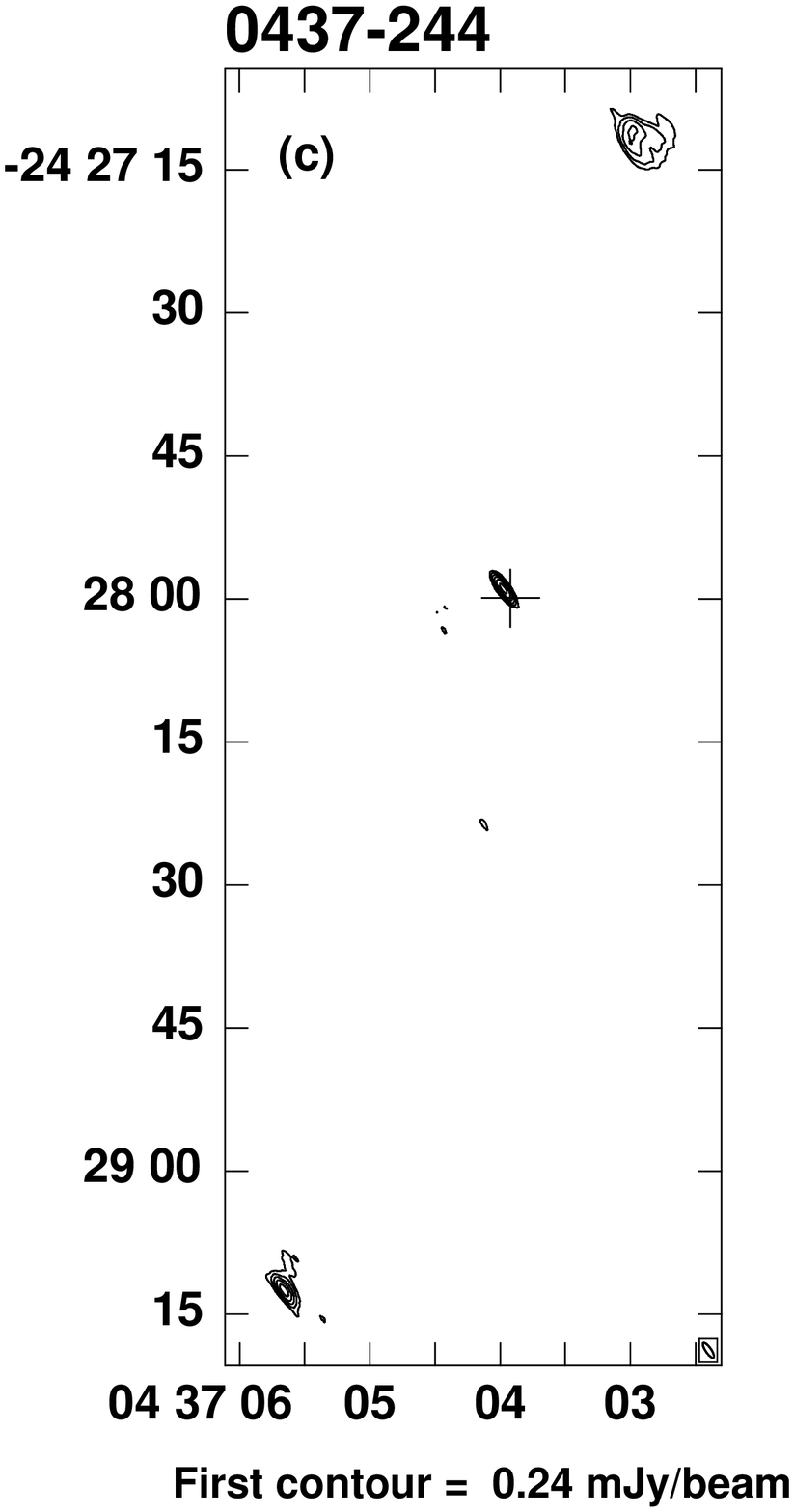,height=4.0in}
}
}
\caption{Radio images of 0437$-$244. (a): total-intensity image at 1365 MHz with an
angular resolution of 4.5$'' \times 3.2''$ along PA = $-70^\circ$;
the spectral index between 1365 and 4935 MHz is superimposed in grey; (b): 
the 15 GHz image with the same resolution as in (a) and (c): the
high-resolution image at 8.4 GHz with an angular resolution of 
1.89$'' \times 0.58''$ along PA $35^\circ$. Contour levels: $-$2, $-$1, 1, 2, 4, 8, 16 ...
$\times$ the first contour which is shown below each image. The cross marks the
position of the optical quasar.}
\end{figure*}

\begin{table} \caption{Observed parameters of the giant quasars}
\begin{tabular}{l l l l l l l}
Source  &  Freq  & rms      &  Comp  & Peak & Int. & $m$ \\
Name    &        & noise    &        &    &      &     \\
        &        &          &        &      &      &     \\
0437$-$244 & 1365 & 100     & N Lobe & 66.2& 281.2&  8.0 \\
           &      &         & S Lobe & 44.1& 165.6&  12.2\\
           &      &         & Core   & 16.9& 18.2 & $<\,0.8$\\
           &      &         & Total  &      & 471  &     \\
           & 1665 & 150     & N Lobe & 53.4& 198.8&  7.7 \\
           &      &         & S Lobe & 35.8& 106.7&  9.2 \\
           &      &         & Core   & 15.2&  16.0&  $<\,1.1$\\
           &      &         & Total  &      &  321 &     \\
           & 4635 & 55      & N Lobe & 22.2& 74.5 &  8.2 \\
           &      &         & S Lobe & 15.9& 42.5 &  10.3\\
           &      &         & Core   & 12.9& 13.6 &  $<\,0.9$\\
           &      &         & Total  &      &  130 &     \\
           & 4935 & 55      & N Lobe & 20.3& 66.6 &  7.8 \\
           &      &         & S Lobe & 14.8& 38.7 &  6.7 \\
           &      &         & Core   & 12.5& 13.0 &  $<\,0.9$\\
           &      &         & Total  &      &  121 &     \\
           & 8450 & 45      & N Lobe &  2.2& 19.2 &  $-$ \\
           &      &         & S Lobe &  5.7& 17.0 &  $-$ \\
           &      &         & Core   & 10.5& 11.6 &  $-$ \\
           &      &         & Total  &      &  46.6&     \\
           &14965 & 170     & N Lobe &  4.2& 15.1 &  $-$ \\
           &      &         & S Lobe &  2.8& 6.63 &  $-$ \\
           &      &         & Core   &  2.7& 4.91 &  $-$ \\
           &      &         & Total  &      & 30.9 &     \\
1025$-$229 & 1365 & 165     & N Lobe & 78.6& 286.5& 9.0 \\
           &      &         & S Lobe & 18.4& 211.1& 10.7\\
           &      &         & Core   & 13.5& 21.8 & $<\,0.7$\\
           &      &         & Total  &      & 560  &     \\
           & 1665 & 150     & N Lobe & 64.3& 214.5& 8.8 \\
           &      &         & S Lobe & 16.6& 155.1& 10.5\\
           &      &         & Core   & 11.0& 16.91& $<\,1.0$\\
           &      &         & Total  &      &  421 &     \\
           & 4635 & 65      & N Lobe & 28.2& 66.9 & 7.1 \\
           &      &         & S Lobe &  6.5& 39.2 & 10.2 \\
           &      &         & Core   & 10.1& 11.2 & $<\,1.0$\\
           &      &         & Total  &      & 128  &     \\
           & 4935 & 70      & N Lobe & 26.1& 59.8 & 8.5 \\
           &      &         & S Lobe & 6.0 & 39.2 & 8.5 \\
           &      &         & Core   & 10.2& 11.2 & $<\,1.0$\\
           &      &         & Total  &      & 118  &     \\
           & 8450 & 45      & N Lobe & 3.2 & 15.9 & $-$ \\
           &      &         & S Lobe & 0.8 & 1.95 & $-$ \\
           &      &         & Core   & 10.0 & 11.03& $-$ \\
           &      &         & Total  &      & 40   &     \\
           &14965 & 230     & N Lobe & 3.8 & 8.23 & $-$ \\
           &      &         & S Lobe & 3.8 & 17.8 & $-$ \\
           &      &         & Core   & 8.7 & 10.0 & $-$ \\
           &      &         & Total  &      & 36   &     \\
\end{tabular}
Note:  Frequency is expressed in MHz, rms in units of $\mu$Jy/beam, peak 
flux densities are in units of mJy/beam, integrated flux density is in mJy and 
$m$, is the scalar percentage polarization.
\end{table}

The GQ 0437$-$244 is a  classical double-lobed FRII radio source with well-defined hotspots at the
outer edges, a core contributing about 10 per cent of the total flux density at an 
emitted frequency of 8 GHz, and signs of a possible jet towards the southern component.
The spectral indices, $\alpha$ (defined as S $\propto \nu^{-\alpha}$),
at the peaks of emission in the outer lobes are about 0.6 and 0.9
for the northern and southern components respectively. In both components the spectral
indices steepen significantly up to distances of about 20 arc sec (about 170 kpc) from the peaks of 
emission (Figure 2). The errors in the spectral indices have been estimated assuming 
an error of 3 per cent in the measured flux density. 
Adopting the formalism of Myers \& Spangler (1985), the age estimates
due to synchrotron radiative losses in the Kardashev-Pacholczyk model (Pacholczyk 1977) 
are 5.8$\times10^{7}$ and 2.7$\times10^{7}$ yr for the northern and 
southern lobes respectively. The injection spectra have been estimated from the 
spectral indices of the hotspots, which have been defined to be the 
peaks of emission in the outer lobes.
The integrated spectra of the northern and southern lobes are straight between 1.4 
and 15 GHz with
spectral indices of 1.16$\pm$0.07  and 1.10$\pm$0.09  respectively since these
are dominated by the hotspots. 
The core has a flat spectrum between 1.4 and 8.4 GHz with a spectral index 
of 0.23$\pm$0.10 but shows evidence of steepening
between 8.4 and 15 GHz, with a spectral index of 1.50$\pm$0.24. Such a steep spectral
index is uncommon; this is the only
quasar in our sample of Molonglo radio sources (cf. IC98)
with a core spectral index between 8.4 and 15 GHz which is steeper than about 0.5.
The steep core spectrum is possibly due to variability of the core flux density. The integrated 
spectrum of the entire source between  408 MHz and  15 GHz 
is 0.94$\pm$0.05.

\begin{figure}
\vbox{
\vspace{-1.0 in}
\psfig{figure=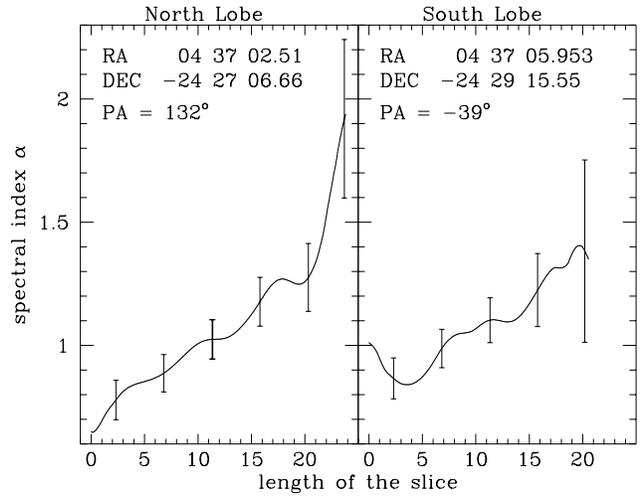,height=3.5in}
}
\caption{The variation in spectral index between 1365 and 4935 MHz for the northern
and southern lobes of 0437$-$244 along the position angles of the slices indicated in
the panels above. The positions of the origins in the lobes are also listed above. 
The error bars indicate $\pm1\sigma$ errors to the spectral indices. }
\end{figure}

\begin{figure*}
\vbox{
\hbox{
\hspace{1.0 in}
\psfig{figure=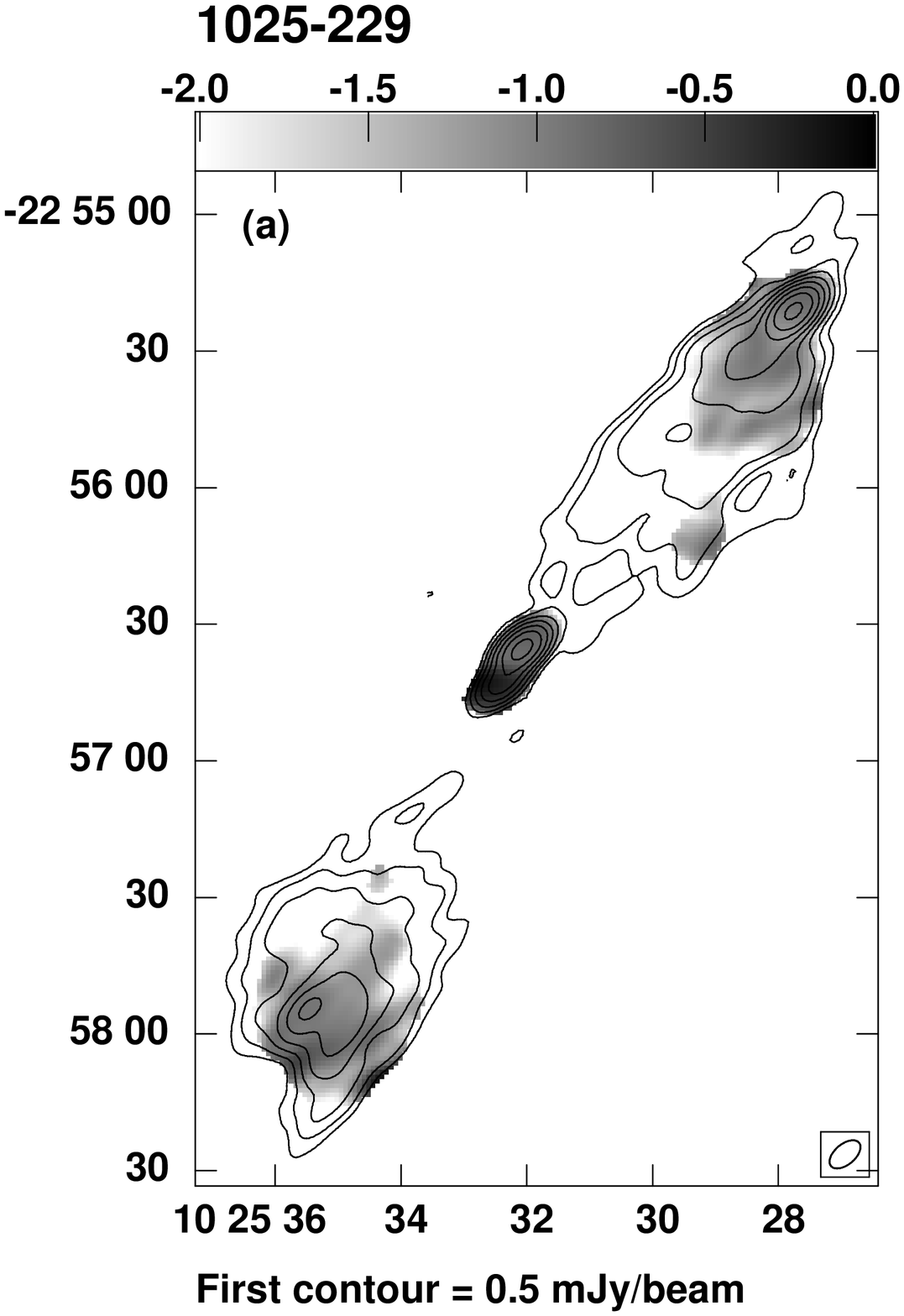,height=3.6in}
\hspace{0.5 in}
\psfig{figure=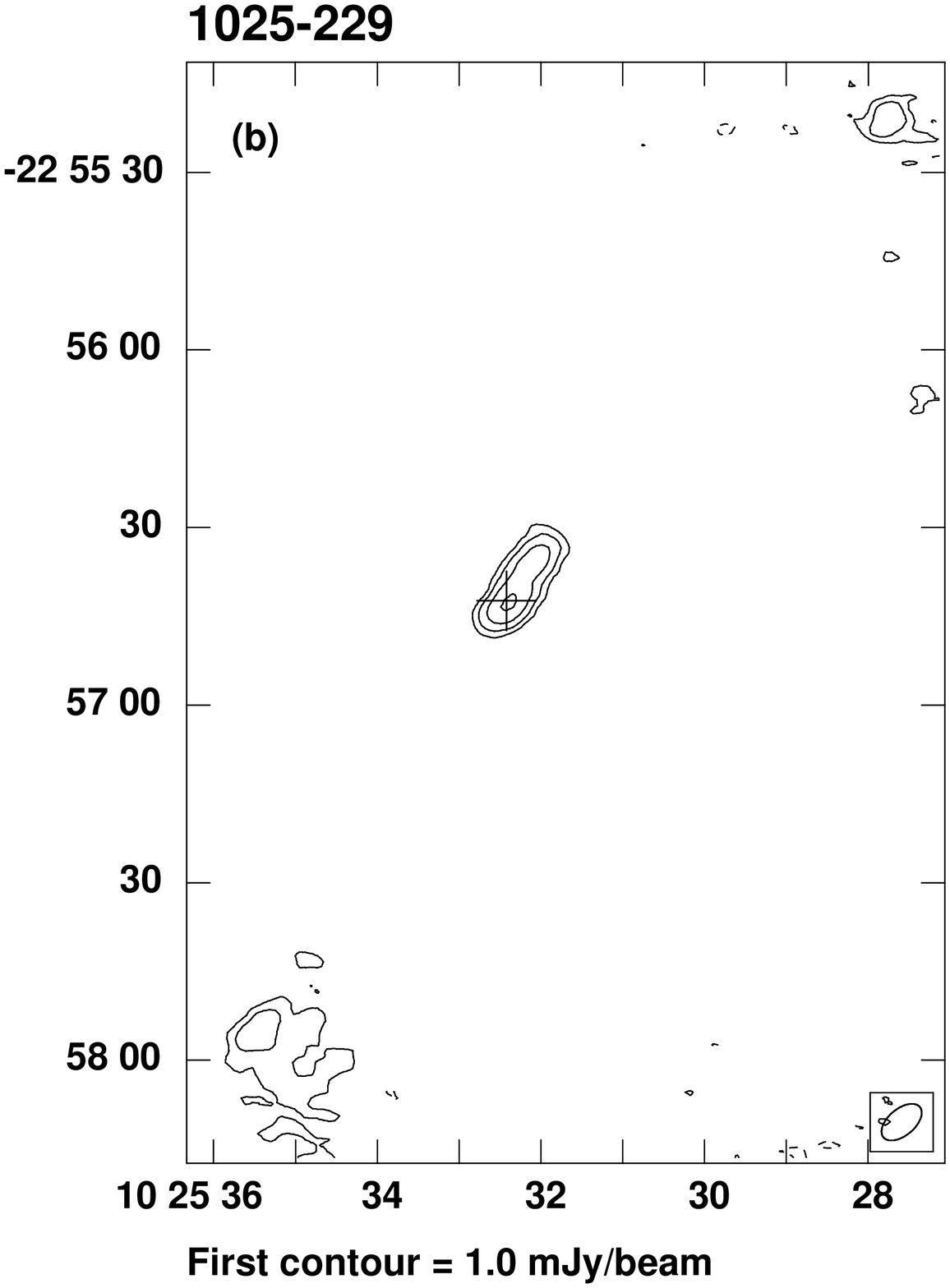,height=3.5in}
}
\hbox{
\hspace{0.75 in}
\psfig{figure=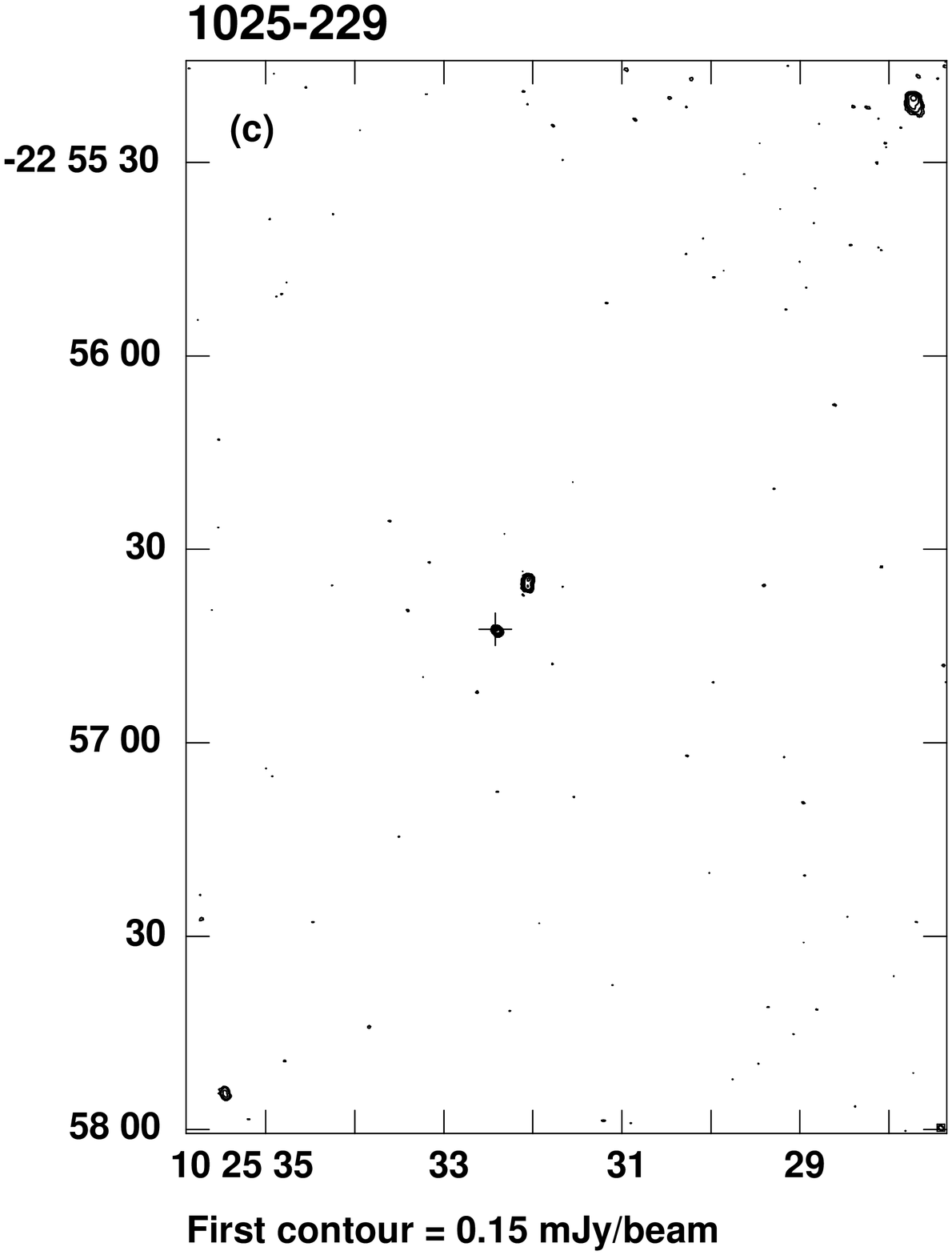,height=3.5in}
\hspace{0.5 in}
\psfig{figure=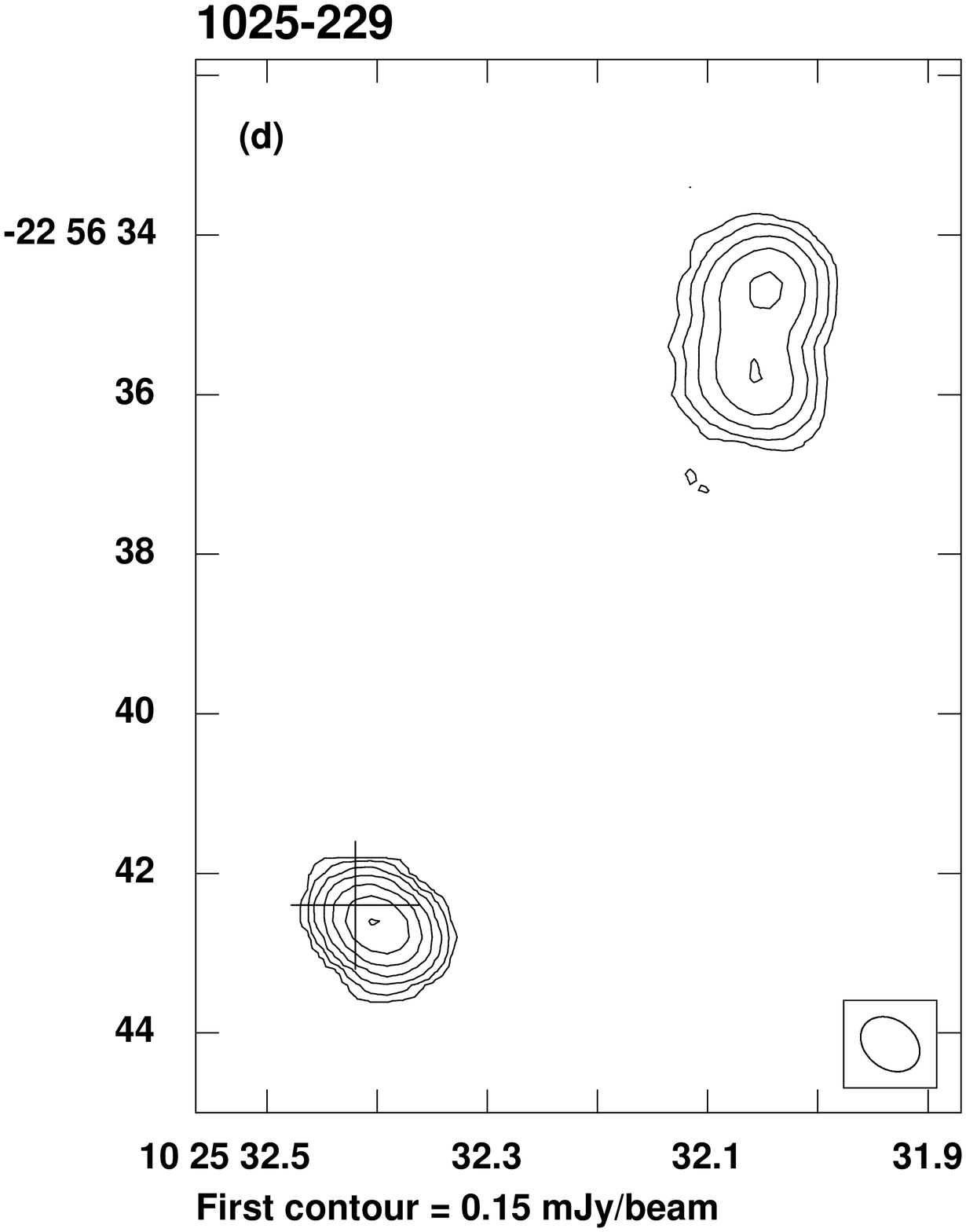,height=3.5in}
}
}
\caption{Radio images of 1025$-$229. (a): total-intensity image at 1365 MHz
with an angular resolution of  8.0$'' \times 4.5''$ along PA $-50^\circ$; 
the spectral index between 1365 and 4935 MHz is superimposed in grey; (b):
the 15 GHz image with the same angular resolution as in (a) and (c): the
high-resolution image at 8.4 GHz of the entire source with an angular 
resolution of 0.80$'' \times 0.61''$ along PA $53^\circ$; (d): the 
high-resolution 8.4 GHz  image of the core and the component close to it. 
Contour levels: $-$2, $-$1, 1, 2, 4, 8, 16 ...
$\times$ the first contour which is shown below each image. The cross marks the
position of the optical quasar.}
\end{figure*}

We have computed the rotation measures for both lobes of the giant quasars
using two widely spaced L band (1365 MHz and 1665 MHz) and the C band
(4935 MHz) frequencies. For 0437$-$244, the northern lobe has an 
integrated RM of about 13.6$\pm$1.3
rad m$^{-2}$, while for the southern lobe the RM is about 4.2$\pm$3.4 rad m$^{-2}$.
The depolarization parameter,
defined to be the ratio of the degree of polarization at 1365 MHz to that at 4935 MHz
is close to about 1, indicating that there is no significant depolarization
till about 1365 MHz (cf. IC98).

\subsubsection{1025$-$229}
This is also a well-defined FRII radio source with two hotspots in
the southern lobe, a core contributing about 12 per cent of the total
flux density of the source, and a jet-like structure close to the radio 
core (Figures 3a-d). The high-resolution 8.4 GHz image of the core region
resolves the core and the jet-like structure; the latter consists of two 
components along a PA of $-7^\circ$. This is significantly different from
the PA of $-38^\circ$ defined by the core and the northern hotspot 
(Figures 3c and d). A deep optical image might help clarify whether this
feature is an unrelated source. However, the similarity in RM of the jet-like
feature and the lobes, which are listed later in this section, 
suggests that these might be related. The spectral index 
slices for the two lobes are presented in 
Figure 4. The spectral index of the northern hotspot is about 0.9 and
steepens to about 1.5 over a distance of 40$''$ (about 220 kpc), while
for the southern lobe both the peaks of emission have a spectral index
of 0.6 which steepens to about 1.7 over a distance of about 40$''$.
Adopting the formalism of Myers \& Spangler (1985), the age estimates
due to synchrotron radiative losses in the Kardashev-Pacholczyk model (Pacholczyk 1977)
are 7.8$\times10^{7}$ and 1.2$\times10^{8}$ yr for the northern and southern lobes 
respectively. The injection spectra have again been estimated from the
hotspot spectral indices. Estimating the ages for a subset of 10 sources
chosen at random from our Molonglo sample (IC98) yields
ages in the range of 0.6$-6.5\times10^7$ yr with a median value of about 2.3$\times10^7$ yr.
The integrated spectral index of the northern and southern lobes between
1365 MHz and 15 GHz are 1.31$\pm$0.09 and 1.08$\pm$0.06 respectively.
The spectral index of the entire source is 0.90$\pm$0.04  between 408 MHz and 15 GHz,
while that of the core is 0.26$\pm$0.10  between 1.665 and 15 GHz. 
The depolarization parameter
between 1365 and 4935 MHz is again close to about 1, while the
rotation measures estimated from the two L-band (1365 MHz and 1665 MHz)
and C-band data are $-$21.3$\pm$2.3, $-$15.3$\pm$0.9 and $-$21.5$\pm$3.6 
rad m$^{-2}$ for the northern
and southern lobes and the steep-spectrum jet-like feature close to
the radio core.

\begin{figure}
\vbox{
\vspace{-1.0 in}
\psfig{figure=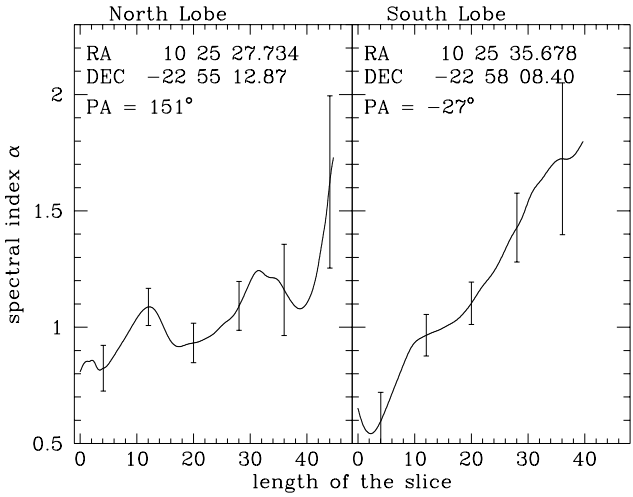,height=3.5in}
}
\caption{The variation in spectral index between 1365 and 4935 MHz for the northern
and southern lobes of 1025$-$229 along the position angles of the slices indicated in
the panels above. The positions of the origins in the lobes are also listed above.
The error bars indicate $\pm1\sigma$ errors to the spectral indices.
}
\end{figure}

\section{The sample of giant radio sources}
To understand the evolution of GRSs, examine their consistency with
the unified schemes and use them as probes of their environment
and the intergalactic medium, we have compiled a sample of known
giant radio sources from the literature. A GRS is defined to be one with
an overall projected linear size $\geq$1 Mpc. Our sample of known GRSs consists
of 53 sources, 48 of which are associated with galaxies and the remaining 5
with quasars. These are listed in Table 3, which is arranged as
follows. Columns 1 and 2: source name and an alternative name. Most sources 
have peaks of emission in the outer parts of the lobes and either belong to FRII or
are in FRI/FRII transition region. Those with a clear FRI structure such as in
3C129 and 3C130 are marked with an asterisk in column 1.
Column 3:
optical identification where G and Q denote galaxy and quasar respectively;
column 4: redshift; columns 5 and 6: the largest angular size in arcsec and 
the corresponding projected linear size in Mpc; 
column 7: the luminosity at 1.4 GHz in units of W Hz$^{-1}$;
column 8: the fraction of emission from the core, f$_c$,  at an emitted 
frequency of 8 GHz, with a $\sim$ sign indicating those sources whose core
flux density has been estimated by us from the available images; 
columns 9 and 10: the 
minimum energy density, u$_{min}$, in units of 10$^{-14}$ J m$^{-3}$ and the 
equipartition magnetic field, B$_{eq}$, in nT (1T = 10$^{4}$ G); 
column 11: the ratio, r$_{\theta}$ defined to be $>$1, of the separation of the 
oppositely directed components from the nucleus; column 12: the misalignment
angle, $\Delta$, defined to be the supplement of the angle formed at the nucleus
by the outer hotspots; column 13: references for the radio structure.
The minimum energy density and equipartition magnetic field have been estimated
(Miley 1980; Longair 1994) for the extended emission assuming a cylindrical 
geometry, a filling factor of unity and that
energy is distributed equally between relativistic electrons and protons. 
The size of the lobes has been estimated from the lowest contours in the 
available images, and the core flux density has been subtracted while estimating
the minimum energy density. In the case of a few sources, such as the 7C ones, for which 
reliable lobe sizes could not be obtained, u$_{min}$ and B$_{eq}$ have not been
listed in the Table.
The luminosity has been estimated between 10 MHz
and 10 GHz using the known spectral indices of the extended emission. For
estimating the error in u$_{min}$ we have assumed errors of 5 per cent in
the flux density, 0.1 in spectral index and 25 per cent in the volume.

\begin{table*} \caption {The sample of giant radio sources }
\begin{tabular}{llllrlllrllrl}
Source     &Other  &Id & z      & LAS  & LLS  &P$_{1.4}$& f$_c$ &u$_{min}$&B$_{eq}$&$r_{\theta}$&$\Delta$ & Ref\\
Name       &Name   &   &        &$''$  & Mpc  &WHz$^{-1}$&      &Jm$^{-3}$&  nT   &      & $^\circ$ &    \\
           &       &   &        &      &      &       &         &         &       &      &     &       \\
0017$-$205 &MRC    & G & 0.197  &  372 & 1.55 & 26.20 & 0.063   &    $-$  &  $-$    & 1.41 &  3  & 1   \\
0055$+$300 &NGC315 & G & 0.0167 & 3480 & 1.64 & 24.82 & 0.32    &   1.94  &  0.137  & 2.04 &  9  & 2,3 \\   
0109$+$492 &3C35   & G & 0.067  &  635 & 1.10 & 25.64 & 0.035   &   2.46  &  0.154  & 1.06 &  5  & 4   \\   
0114$-$476 &PKS    & G & 0.146  &  702 & 2.35 & 26.51 &$\sim$ 0.012   &   4.25  &  0.202  & 1.39 & 18  & 5   \\   
0132$+$376 &3C46   & G & 0.4373 &  163 & 1.09 & 27.21 & 0.0078  &  38.74  &  0.611  & 1.56 &  7  & 6   \\   
0136$+$397 &4C39.04& G & 0.2107 &  343 & 1.50 & 26.30 & 0.016   &   4.49  &  0.208  & 1.19 &  2  & 7,8 \\   
0157$+$405 &4C40.08& G & 0.078  &  840 & 1.67 & 25.62 & 0.018   &   0.99  &  0.098  & 2.54 &  0  & 8,9 \\   
0211$-$479 &PKS    & G & 0.2195 &  378 & 1.70 & 26.54 &$\sim$ 0.0069  &   6.68  &  0.254  & 1.15 &  1  & 5   \\   
0309$+$411 &B3     & G & 0.136  &  570 & 1.80 & 26.01 & 0.54    &   0.97  &  0.097  & 1.67 & 14  & 10  \\   
0313$+$683 &WENSS  & G & 0.0902 &  894 & 2.01 & 25.64 & 0.19    &   4.46  &  0.207  & 1.45 &  6  & 11  \\   
0313$-$271 &MRC    & G & 0.216  &  227 & 1.01 & 26.06 & 0.011   &   3.31  &  0.179  & 1.25 & 10  & 1   \\   
0319$-$454 &PKS    & G & 0.0633 & 1538 & 2.54 & 25.81 & 0.023   &   1.57  &  0.123  & 2.22 &  1  & 12  \\   
0424$-$728 &PKS    & G & 0.1921 &  346 & 1.42 & 26.51 &$\sim$ 0.039   &  12.41  &  0.346  & 1.09 &  6  & 5   \\   
0437$-$244 &MRC    & Q & 0.84   &  128 & 1.06 & 27.46 & 0.10    &  47.50  &  0.677  & 1.54 &  0  & 13  \\   
0445$+$449$^*$ &3C129  & G & 0.021  & 1800 & 1.06 & 25.21 & 0.023   &   6.53  &  0.251  & $-$  & $-$ & 4   \\   
0448$+$519$^*$ &3C130  & G & 0.109  &  584 & 1.54 & 26.29 & 0.043   &   5.22  &  0.224  & $-$  & $-$ & 4   \\   
0503$-$286 &MRC    & G & 0.038  & 2400 & 2.48 & 25.26 &$\sim$ 0.029   &   1.28  &  0.111  & 1.89 & 14  & 14,15\\  
0511$-$305 &PMN    & G & 0.0583 &  684 & 1.05 & 25.70 & 0.017   &   3.93  &  0.195  & 1.88 & 18  & 5   \\   
0634$-$205 &PMN    & G & 0.056  &  810 & 1.20 & 26.07 & $<$\,0.021   &   7.17  &  0.263  & 1.32 &  2  & 16,17\\  
0654$+$482 &7C     & G & 0.776  &  135 & 1.10 & 26.69 &  $-$    &    $-$  &  $-$    & $-$  & $-$ & 18  \\
0707$-$359 &PKS    & G & 0.2182 &  492 & 2.21 & 26.71 &$\sim$ 0.10    &   8.03  &  0.278  & 1.44 &  7  & 5   \\   
0744$+$558 &DA240  & G & 0.0356 & 2040 & 1.99 & 25.50 & 0.077   &   0.61  &  0.077  & 1.39 &  2  & 19,20\\  
0821$+$695 &8C     & G & 0.538  &  402 & 2.94 & 26.29 & 0.22    &   3.23  &  0.177  & 1.12 &  1  & 21   \\  
0915$+$320$^*$ &B2     & G & 0.062  &  660 & 1.07 & 24.69 & 0.14    &   2.12  &  0.143  & $-$  & $-$ & 22   \\  
0945$+$734 &4C73.08& G & 0.0581 &  884 & 1.35 & 25.77 & 0.024   &   1.68  &  0.127  & 1.70 &  6  & 4    \\  
1003$+$351 &3C236  & G & 0.0988 & 2340 & 5.70 & 26.41 & 0.46    &   1.32  &  0.113  & 1.68 &  2  & 19,23,24\\
1025$-$229 &MRC    & Q & 0.309  &  198 & 1.11 & 26.54 & 0.12    &  16.73  &  0.402  & 1.25 &  8  & 13   \\  
1029$+$571$^*$ &HB13   & G & 0.034  & 1110 & 1.03 & 24.50 & 0.19    &   4.30  &  0.204  & $-$  & $-$ & 25   \\  
1058$+$368 &7C     & G & 0.75   &  158 & 1.22 & 26.86 &  $-$    &    $-$  &  $-$    & $-$  & $-$ & 18   \\
1127$-$130 &PKS    & Q & 0.6337 &  297 & 2.30 & 27.53 & 0.087   &   6.93  &  0.259  & 1.21 &  0  & 26   \\  
1144$+$352 &WENSS  & G & 0.063  &  701 & 1.15 & 25.09 & 0.95    &   1.18  &  0.107  & 1.20 &  10 & 27   \\
1158$+$351 &87GB   & G & 0.55   &  140 & 1.03 & 26.79 & 0.015   &  33.22  &  0.566  & 1.24 &  5  & 28   \\  
           &       &   &        &      &      &       &         &         &       &      &     &       \\
1209$+$745 &4C74.17& G & 0.107  &  420 & 1.09 & 25.52 & 0.089   &   2.59  &  0.158  & 1.55 & 18  & 29   \\  
1218$+$639 &TXS    & G & 0.2    &  420 & 1.77 & 26.69 &  $-$    &    $-$  &  $-$    & $-$  & $-$ & 30   \\
1232$+$216 &3C274.1& G & 0.422  &  152 & 1.00 & 27.32 & 0.031   &  47.53  &  0.677  & 1.21 &  6  & 31   \\  
1312$+$698 &DA340  & G & 0.106  &  420 & 1.09 & 25.93 &  $-$    &    $-$  &  $-$    & $-$  & $-$ & 30   \\
1331$-$099 &PKS    & G & 0.081  &  820 & 1.68 & 26.06 & 0.11    &   4.84  &  0.216  & 1.09 &  3  & 32   \\  
1349$+$647 &3C292  & G & 0.71   &  133 & 1.06 & 28.09 & 0.0013  & 162.53  &  1.252  & 1.12 &  3  & 31   \\  
1358$+$305 &B2     & G & 0.206  &  649 & 2.80 & 26.03 & 0.011   &   1.17  &  0.106  & 2.13 &  5  & 33   \\  
1452$-$517 &MRC    & G & 0.08   & 1218 & 2.48 & 25.66 & 0.35    &   1.10  &  0.103  & 1.60 & 12  & 8,34 \\  
1519$+$513 &87GB   & G & 0.37   &  258 & 1.59 & 27.06 & 0.017   &  18.20  &  0.419  & 1.08 &  0  & 28   \\  
1545$-$321 &PKS    & G & 0.1085 &  498 & 1.31 & 26.03 & $<$\,0.032   &  12.29  &  0.344  & 1.02 &  4  & 5    \\  
1549$+$202 &3C326  & G & 0.0895 & 1173 & 2.63 & 26.15 & 0.012   &   2.00  &  0.139  & 1.95 &  3  & 35   \\  
1602$+$376 &7C     & G & 0.814  &  175 & 1.44 & 26.93 &  $-$    &    $-$  &  $-$    & $-$  & $-$ & 18   \\
1626$+$518 &WENSS  & G & 0.056  & 1140 & 1.68 & 25.10 & 0.14    &    $-$  &  $-$    & 1.44 & 5   & 36   \\
1636$+$418 &7C     & G & 0.867  &  130 & 1.09 & 26.67 &  $-$    &    $-$  &  $-$    & $-$  & $-$ & 18   \\
1637$+$826 &NGC6251& G & 0.023  & 3120 & 2.00 & 24.66 & 0.75    &   0.38  &  0.061  & 1.71 & 26  & 37,38\\  
1701$+$423 &7C     & G & 0.476  &  180 & 1.24 & 26.47 &  $-$    &    $-$  &  $-$    & $-$  & $-$ & 18   \\
1721$+$343 &4C34.47& Q & 0.2055 &  244 & 1.05 & 26.51 & 0.60    &   8.02  &  0.278  & 1.05 & 1   & 39,40\\  
1834$+$620 &WENSS  & G & 0.519  &  197 & 1.42 & 27.19 & $-$     & 20.85   &  0.449  &1.03  & 0   & 41 \\
1910$-$800 &PKS    & G & 0.346  &  366 & 2.18 & 26.79 &$\sim$ 0.059   &  10.16  &  0.313  & 1.06 &  0  & 5    \\  
2043$+$749 &4C74.26& Q & 0.104  &  610 & 1.55 & 26.03 & 0.51    &   3.02  &  0.171  & 1.04 &  7  & 42   \\  
2309$+$184 &3C457  & G & 0.427  &  205 & 1.36 & 27.38 & 0.013   &  32.96  &  0.564  & 1.00 &  6  & 43   \\  
\end{tabular}

\vspace*{1ex}
\begin{flushleft}
References: 1. Kapahi et al. 1998b; 2. Bridle et al. 1976; 3. Willis et al. 1981; 
4. J\"{a}gers 1986; 5. Subrahmanyan, Saripalli \& Hunstead 1996; 6. Gregorini et al. 1988;
7. Hine 1979; 8. Saripalli 1988; 9. Vigotti et al. 1989 ; 10. de Bruyn 1989; 
11. Schoenmakers et al. 1998a; 12. Saripalli, Subrahmanyan \& Hunstead 1994; 
13. Present work; 14. Saripalli et al. 1986; 15. Subrahmanya \& Hunstead 1986; 
16. Danziger, Goss \& Frater 1978; 17. Kronberg, Wielebinski \& Graham 1986;  
18. Cotter, Rawlings \& Saunders 1996; 19. Willis, Strom \& Wilson 1974; 
20. Strom, Baker \& Willis 1981; 21. Lacy et al. 1993; 22. Ekers et al. 1981;  
23. Strom \& Willis 1980; 24. Barthel et al. 1985; 25. Masson 1979; 
26. Bhatnagar, Gopal-Krishna \& Wisotzki 1998;  27. Schoenmakers et al. 1998b;
28. Machalski \& Condon, 1985; 
29. van Breugel \& Willis  1981;  30. Saunders, Baldwin \& Warner 1987; 
31. Alexander \& Leahy  1987; 32. Saripalli et al. 1996; 33. Parma et al 1996; 
34. Jones \& McAdam 1992; 35. Willis \& Strom  1978; 36. R\"ottgering et al. 1996;  
37. Waggett, Warner \& Baldwin 1977;  38. Willis et al. 1982 ; 
39. J\"{a}gers et al. 1982; 40. Barthel 1987; 41. de Bruyn et al. 1998; 
42. Riley et al 1989; 43. Leahy \& Perley 1991.
\end{flushleft}
\end{table*}

\begin{figure}
\vbox{
\vspace{-1.0 in}
\psfig{figure=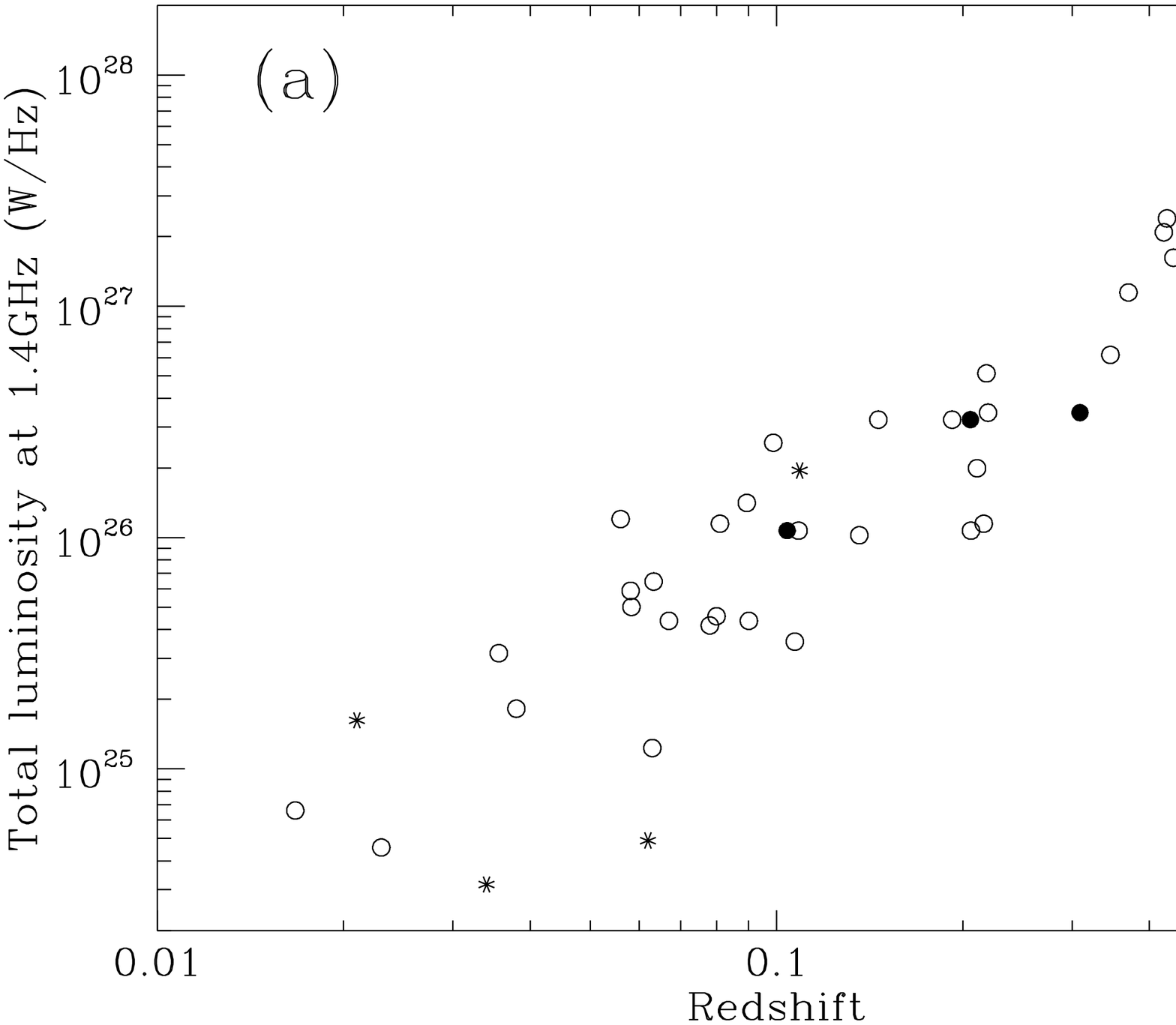,height=3.5in}
\vspace{-0.75 in}
\psfig{figure=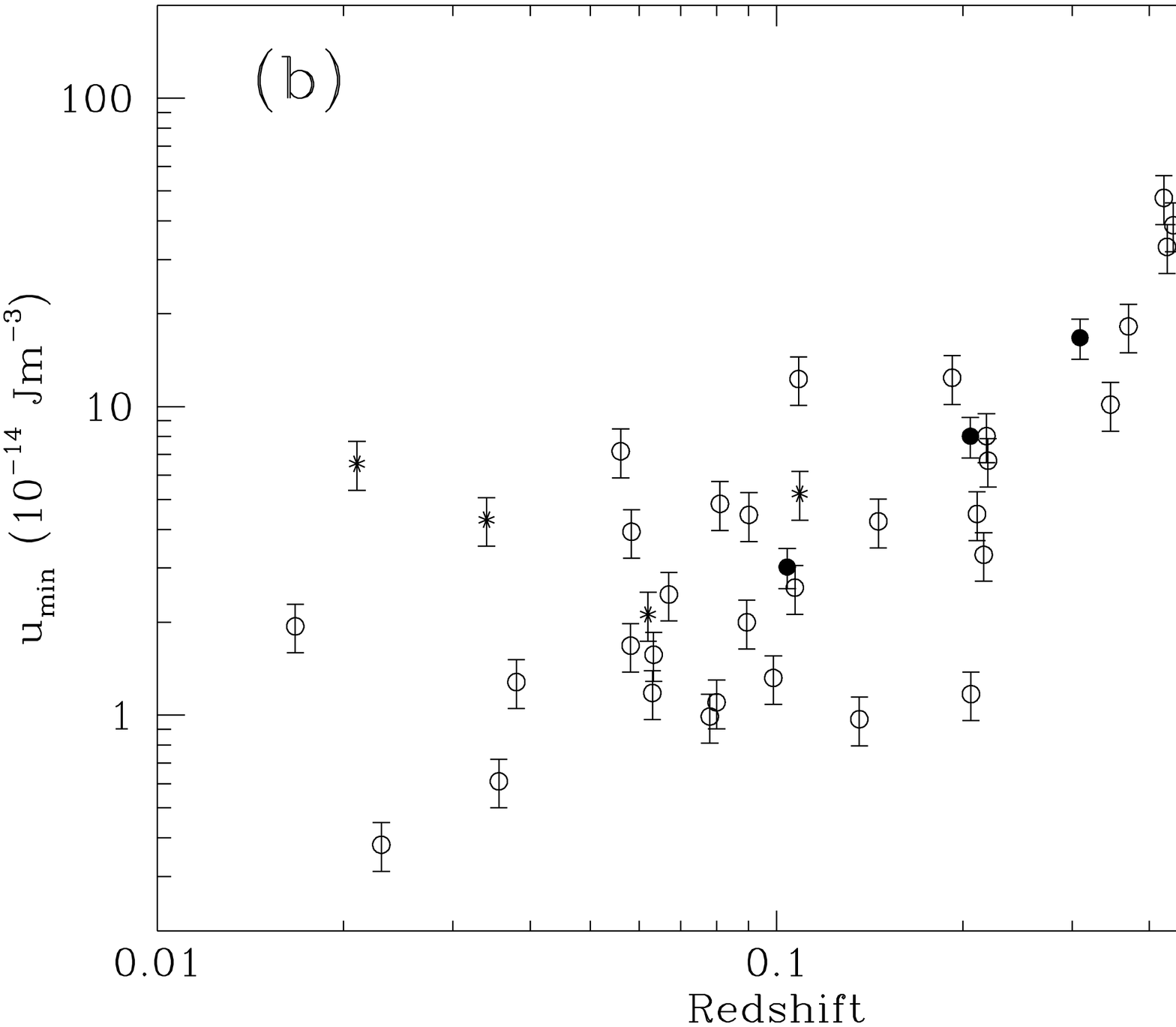,height=3.5in}
}
\caption{{\bf (a)} The total radio luminosity at 1.4 GHz and {\bf (b)} 
the minimum energy density of the lobes against the redshift of the 
giant radio sources. Filled circles denote quasars while open circles denote
all radio galaxies except those with a clear FRI structure, which have been
marked with an asterisk.  
}
\end{figure}

\subsection{Energy density and magnetic field}
The minimum energy density is in the range
of 0.4 to 162 $\times$ 10$^{-14}$ J m$^{-3}$ with a median value of about
4.4 $\times$ 10$^{-14}$ J m$^{-3}$, while the equipartition magnetic
field for the lobes range from 0.06 to 1.25 nT with a median value of 0.2 nT. 
The minimum energy density might be expected to increase with redshift due
to better confinement by the intergalactic medium, and greater dissipation
of energy by the beams interacting with a denser environment. 
An earlier study by Subrahmanyan \& Saripalli (1993) showed marginal
evidence of an increase in minimum energy density or pressure with redshift. 
However, it is important
to confirm whether such trends might be due to possible selection effects. For example,
if the giant sources are chosen from a given flux-density limited sample,
luminosity and minimum energy density could be strongly correlated
with redshift since most sources are close to the flux density limit of the
survey. Although the giant radio sources in our sample have been selected
from surveys ranging from the 3CR to WENSS (Laing, Riley \& Longair 1983;
Rengelink et al. 1997), the observed range in flux density at say 1.4 GHz 
for almost all the objects for which we have been able to evaluate u$_{min}$ are within 
one order of magnitude, while the luminosity spans about 3 orders of magnitude.
Thus although the luminosity and minimum energy density appear correlated with
redshift (Figures 5a and b) with Spearman rank correlation co-efficients of 0.90
and 0.68 respectively, we need to identify a large number of giant sources from
surveys such as WENSS and NVSS to determine whether there is a genuine trend for
the minimum energy density to increase with redshift.

\begin{figure}
\vbox{
\vspace{-1.0 in}
\psfig{figure=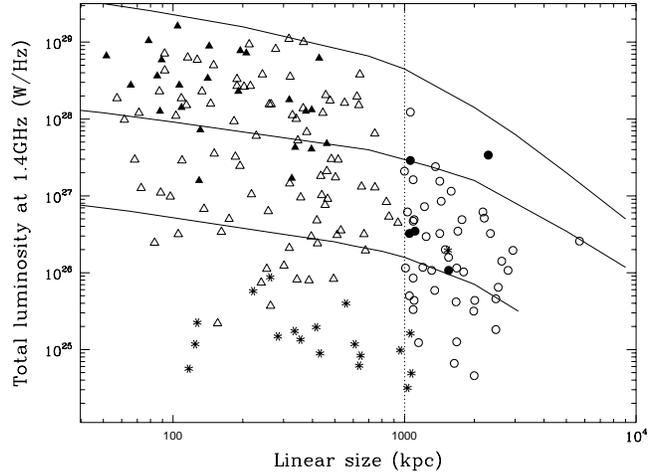,height=3.5in}
}
\caption{The luminosity - linear size or P-D diagram for all 3CR sources with
50 kpc $<$ D $<$ 1 Mpc and our sample of giant sources.
The giant quasars and galaxies are shown by filled and open circles respectively;
while the 3CR quasars and galaxies are shown by filled and open
triangles respectively, except for those with a clear FRI structure. The FRI 
sources are marked with an asterisk. The evolutionary scenarios for sources with jet powers
of 1.3$\times$10$^{40}$,  1.3$\times$10$^{39}$ and 1.3$\times$10$^{38}$ W from
Kaiser et al. (1997) are shown superimposed on the diagram.
}
\end{figure}

\subsection{Evolution of giant radio sources}

We investigate the evolution of the giant radio sources by plotting them in
a power-linear size or P-D diagram along with the complete sample of 3CR
(Laing et al. 1983) radio sources. The P-D diagram is, in principle,
a powerful tool
for investigating the temporal evolution of radio sources (Shklovskii 1963;
Scheuer 1974), 
although in practice the detailed interpretation is debatable
(cf. Baldwin 1982; Kaiser, Dennett-Thorpe \& Alexander 1997; Blundell et al. 1999). 
One must also be careful of possible selection effects since large sources
with weak lobes or cocoons  may sometimes appear to be disconnected and hence
misclassified as independent sources. Also, weak FRI sources with diffuse extended
lobes may have only the bright inner parts detected at large redshifts and hence
appear smaller in size (cf. Neeser et al. 1995). However, all but 4 of the GRSs are
clear FRIIs or are in the FRI/II transition region and have well-defined  peaks of
emission towards their outer edges. Hence their parameters are unlikely to be 
affected by such selection effects. The P-D diagram for the complete sample
of 3CR radio sources with sizes between 50 kpc and 1 Mpc, 
and all the giant sources (Figure 6) show that there is a 
clear deficit of giant sources ($>$1 Mpc) with high radio 
luminosity, suggesting that the luminosity of radio sources decrease as they 
evolve. This trend was suggested earlier (Baldwin 1982; Cotter et al. 1996, 
Kaiser \& Alexander 1997; Kaiser et al. 1997) using small samples of giant
sources, and has now been established with a sample of over 50 giant radio sources.
Our sample of 53 giant sources includes all
those from a complete sample of sources selected from the Molonglo Reference 
Catalogue or MRC (Large et al. 1981; Kapahi et al. 1998a,b and references therein), 
and early results from the
WENSS survey in addition to the searches for giants undertaken by
Cotter et al. (1996) from the 7C survey and Subrahmanyan et al. (1996) from 
the MRC. Given the range of surveys
and the systematic searches for giant sources, the dearth of giant objects
with high radio luminosity is unlikely to be due to any selection effects.
There is also a sharp cutoff in the sizes of the GRSs at about 3 Mpc, with
only one exception, namely 3C236, which has a size of 5.7 Mpc. To investigate
whether there are larger sources which may have been missed, one requires low-frequency
surveys with higher sensitivity to diffuse, low-brightness emission. 
In Figure 6, we superimpose the evolutionary tracks suggested by Kaiser et al. (1997)
for three different jet powers and find that our sample of giant sources is 
roughly consistent with their self-similar models where the lobes lose energy due to
expansion and radiative losses due to both inverse-Compton and synchrotron processes. 
In the models developed by Blundell et al. (1999), the luminosity declines more
rapidly than the Kaiser et al. tracks, and provides a somewhat better fit to the
upper envelope for large linear sizes. The P-D diagram along with these
evolutionary models suggest that the progenitors
of the giants are normal FRII or FRI radio sources, depending on their jet
power, and the giant sources do not represent objects with unusually large
nuclear engines or increased activity in the nucleus.

\begin{figure}
\vbox{
\vspace{-1.0 in}
\psfig{figure=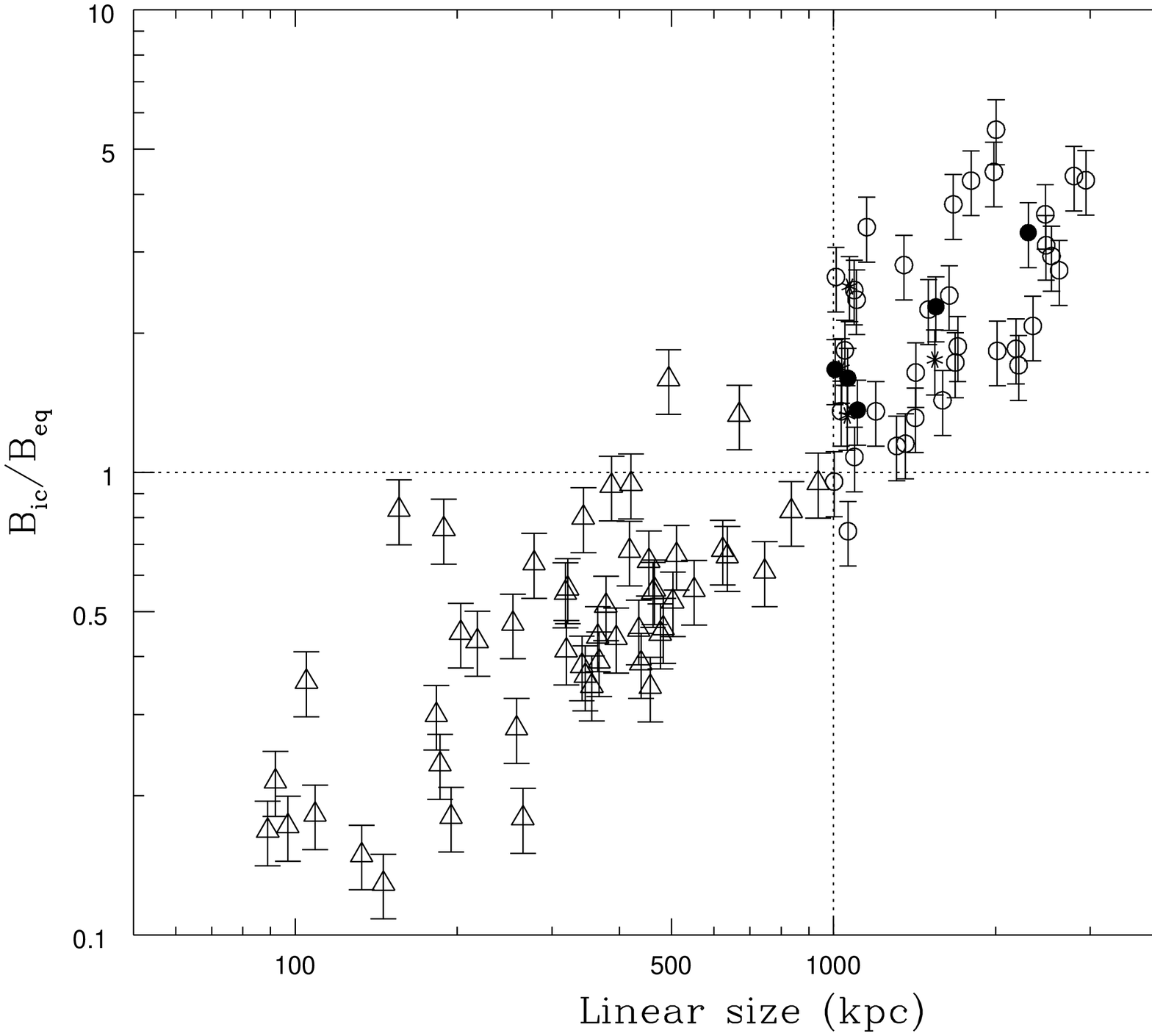,height=3.5in}
\vspace{-0.75 in}
\psfig{figure=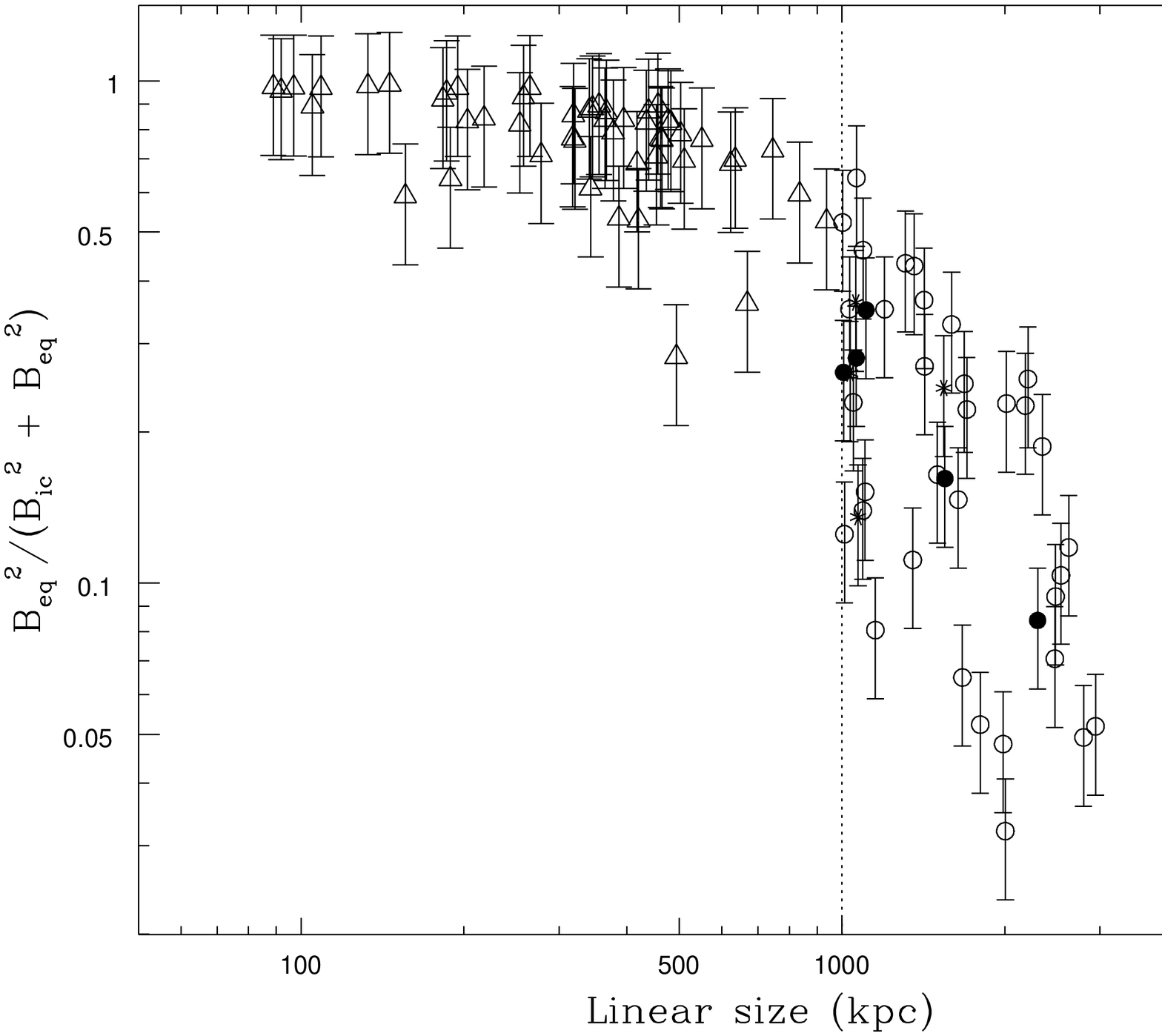,height=3.5in}
\vspace{-0.75 in}
\psfig{figure=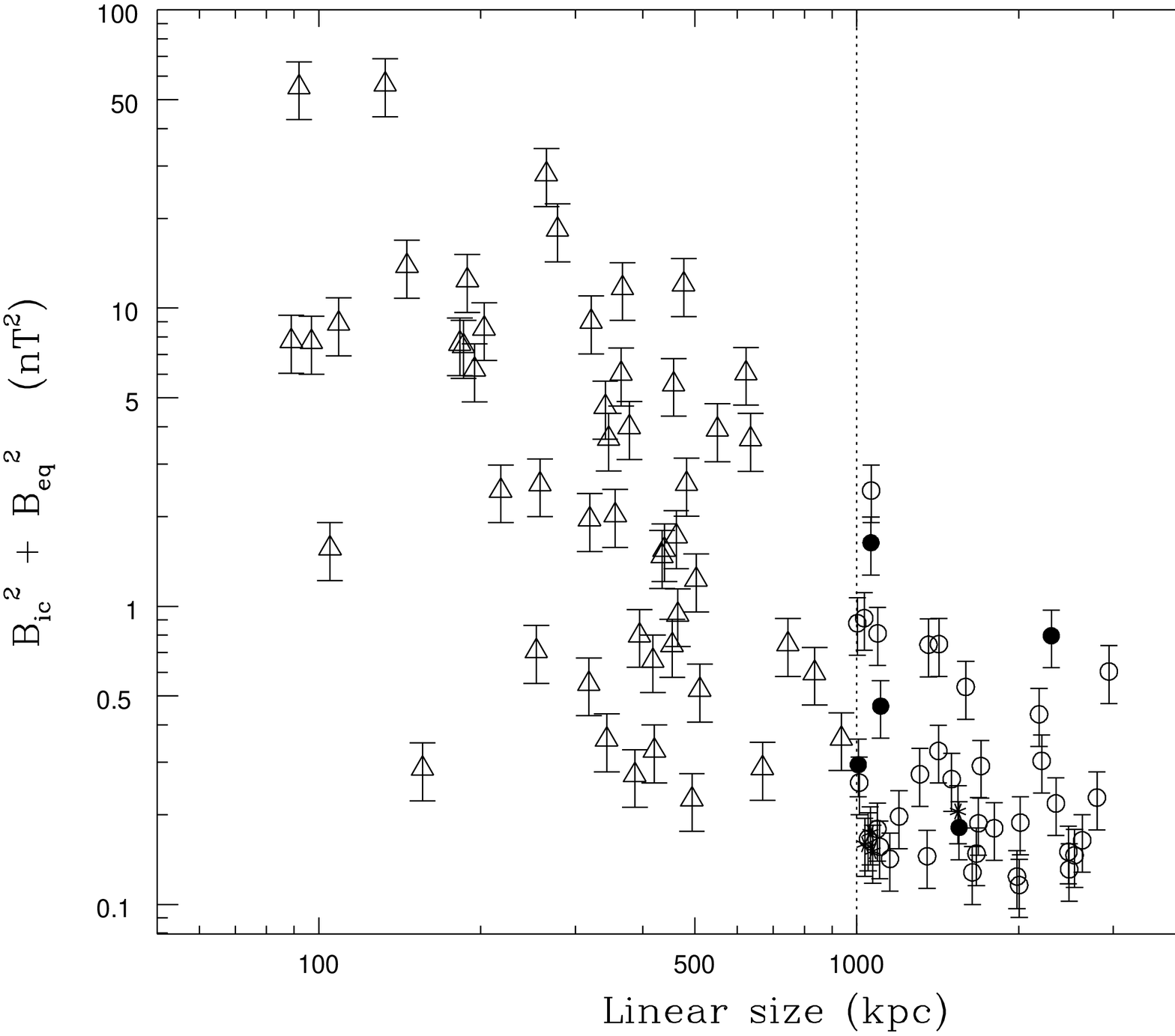,height=3.5in}
}
\caption{{\bf (a)} The ratio  B$_{ic}$/B$_{eq}$; {\bf (b)} 
the ratio  B$_{eq}^2$/(B$_{ic}^2$ + B$_{eq}^2$); {\bf (c)}
 B$_{ic}^2$ + B$_{eq}^2$ plotted against the projected linear 
size of the sample of giant and smaller sources.
In panel (c) the scale of the y axis is in units of nT$^2$, while the x-axis
is in kpc  in all the above panels. The giant quasars and galaxies are shown 
by filled and open circles respectively;
while the 3CR galaxies are shown by open triangles, except for the clear FRI radio
galaxies. The FRI sources are marked with an asterisk.}
\vspace{-0.1 in}
\end{figure}

\subsection{Radiation losses in giant sources}
In this section we investigate the relative importance of synchrotron and
inverse-Compton losses in the evolution of giant radio sources.
The equipartition magnetic field of the lobes for almost all the giant 
sources are less than the equivalent magnetic field of the microwave
background radiation, suggesting that inverse-Compton losses are larger than
the synchrotron radiative losses in the evolution of the lobes of these giant
sources. For comparison, we have computed the equipartition magnetic field
for all the sources, except 3C293 \& 3C321, which are in the 3CR complete sample and have been 
observed by  Leahy \& Williams (1984), Leahy, Muxlow \& Stephens (1989),
Leahy \& Perley (1991), Fernini et al. (1993), Johnson, Leahy \& Garrington (1995),
Fernini, Burns \& Perley (1997) and Hardcastle et al. (1997), 
and also the corresponding inverse Compton field, 
B$_{ic}$=0.32(1 + z)$^2$ nT, at the redshift of the source. The above two
sources were excluded because their bridges have not been well-mapped.
A plot of the linear 
size of the radio source against the ratio of inverse Compton
field to equipartition magnetic field (Figure 7a) shows that synchrotron 
losses dominate over inverse Compton losses for almost all objects below
about a Mpc while the reverse is true for the giant sources. This is 
also illustrated in Figure 7b where we plot the linear size against the
ratio, B$_{eq}^2$/(B$_{ic}^2$ + B$_{eq}^2$), which represents the ratio
of the energy loss by synchrotron radiation to total energy loss due to
both the processes. The ratio is close to about 1
for the small sources, lies between about 0.5 and 1 for sources less than a
Mpc, and falls sharply for the giant sources to a value of about 0.05. 
This is
consistent with similar suggestions made earlier by Gopal-Krishna, Wiita \& Saripalli (1989).
Since the sample of giant radio sources has been compiled from a large number
of surveys complete to different flux density limits, we have checked  and confirmed
these trends by considering 3CR and giant sources over a 
similar luminosity range of 10$^{25}$ to $10^{27}$ W Hz$^{-1}$. This
range was chosen to maximize the number of 3CR and giant radio sources in a 
similar luminosity range. It is also relevant to note here that de Ruiter et al.
(1990) reported a strong trend for decreasing internal energy density with size,
even after taking into account the correlation of size and radio power in their sample.
The energy density of microwave background radiation increases steeply with
redshift, and it is relevant to enquire whether our trends might be due to a 
higher redshift for the giant sources. The  
median redshift for the GRSs  is about 0.15, while for the 
comparison sample it is about 0.26  showing that the importance of inverse-Compton losses
for the giant sources is not due to higher redshifts for these objects. However,
this illustrates that inverse-Compton losses are likely to 
severely constrain the number of GRSs at high redshifts since the 
microwave background energy density increases as $(1 + z)^4$. 

The rate of energy loss  by the electrons due to both inverse Compton 
and synchrotron processes, proportional to (B$_{ic}^2$ + B$_{eq}^2$), is plotted
against the overall linear size for the GRSs and the comparison sample in 
Figure 7c. For the smaller sources where synchrotron losses dominate, the
rate of energy loss decreases with linear size and tapers to a minimum value
of about 0.10  for the GRSs which is set by the equivalent magnetic field of the
microwave background radiation. 
The lifetime of a relativistic electron at an observed frequency, $\nu_o$, 
due to both synchrotron and inverse-Compton losses is given by 
$${\rm \tau =
{5.03\times 10^4 \over [(1+z)\nu_o]^{1/2}B_{eq}^{3/2}[1 + (B_{ic}/B_{eq})^2]} \ \ Myr}, $$
where $\nu_o$ is in MHz and the magnetic fields B$_{ic}$ and B$_{eq}$ are in units of 10$^{-10}$ T.
For sources of $\sim$ 2\,Mpc, the ratio $\rm {B_{ic}/B_{eq}}$
is about 3 (cf. Figure 7a). For the median redshift of the giant galaxies (z $\sim$ 0.1), 
the lifetime of the radiating electron is $\sim 2\times10^8$ yr at 327 MHz. The time scale for
transport of energy from the nucleus is about 7$\times10^7$ yr for a speed of 0.1c.
Thus the source is likely to be visible well after the supply of energy to the outer
lobes has ceased. These sources would be more easily detectable at low radio frequencies, 
and systematic searches for GRSs $>$ 3 Mpc using telescopes such as the GMRT would 
help clarify the late stages of their evolution.

\begin{figure}
\vbox{
\vspace{-1.0 in}
\psfig{figure=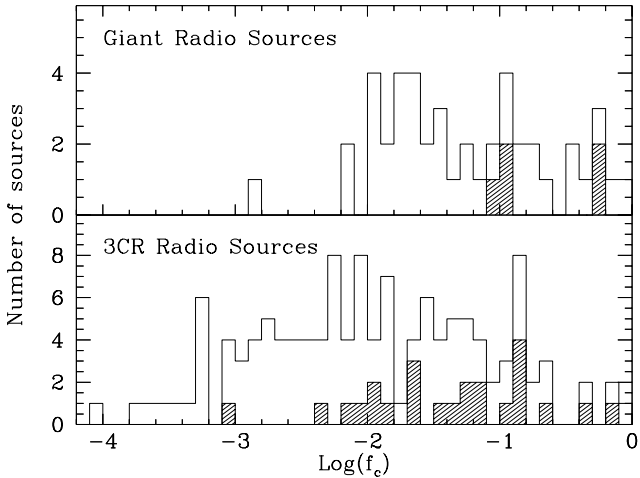,height=3.5in}
}
\caption{The distributions of f$_c$, the fraction of emission from the core at an
emitted frequency of 8 GHz, for the samples of giant sources (upper panel) and
the sample of 3CR sources with 50 kpc $<$ D  $<$ 1 Mpc (lower panel). The quasars
are shown hatched in both panels. 
}
\end{figure}

\section{Constraints on orientation  and environment}
To examine further whether the 
giant radio sources might evolve to large sizes because of a powerful nuclear
engine, we investigate the relative strength of the cores as an indicator of
nuclear activity. Although the degree of core prominence may vary over the 
lifetime of the radio source, it might be possible to arrive at statistically
meaningful results. However, one must also bear in mind that in the unified schemes 
the strength of the core is a statistical indicator of orientation of the
source axis to the
line of sight. Here, the sources at small angles to the line of sight have
prominent cores compared to those at large angles due to the effects of
relativistic beaming (cf. Blandford \& K\"{o}nigl 1979). In the unified
schemes, the radio galaxies and quasars are believed to be intrinsically
similar objects, but appearing to be different because of their different
angles of inclination to the line of sight with the quasars being seen at 
small angles while the radio galaxies lie close to the plane of the sky 
(Scheuer 1987; Barthel 1989; Antonucci 1993; Urry \& Padovani 1995). The
core flux density has been estimated from images with angular resolutions of
about an arcsec or so, and any possible contamination of core flux density
by small-scale jets in the intermediate luminosity objects is minimal and
unlikely to affect the trends reported here.

\subsection{Core prominence}
There have been suggestions that the giant sources have stronger nuclear
activity represented by more prominent cores than the smaller sources 
(e.g. Gopal-Krishna et al. 1989). In this section, we investigate this aspect
using our sample of giant and smaller 3CR sources.
In Figure 8 we show the distributions of the fraction of emission from the
core at an emitted frequency of 8 GHz, f$_c$,  for all the giant sources (upper panel)
and the complete sample of 3CR sources with sizes between 50 kpc and 1 Mpc (lower panel). The
quasars are shown hatched. For both the giant sources as well as the 3CR sources
it is clear that the quasars have more prominent cores than the radio galaxies,
suggesting that the giants associated with quasars are also at smaller angles
to the line-of-sight than the giant radio galaxies, consistent with the unified
scheme. The median values of f$_c$ for the giant quasars
and radio galaxies are 0.12 and 0.034 respectively. However, for the 3CR sources
in the size range of 50 kpc to 1 Mpc, the median values of f$_c$ are 0.052
and 0.0071 for the quasars and radio galaxies respectively. At first glance it 
appears that the giant galaxies do have more prominent cores than the smaller 3CR 
galaxies, and one might be tempted to suggest that they have stronger nuclear engines. The
difference for the quasars needs to be examined further since there are only 5 
giant quasars.
Since many of the giant galaxies are in the borderline of the FRI/FRII classification,
we examine the dependence of f$_c$ on the total radio luminosity
for the complete sample of 3CR radio sources with sizes between 50 kpc and 1 Mpc, 
and our sample of giant sources (Figure 9a).
There is a clear tendency of the weaker radio sources to have more prominent cores.
While the fading of lobes in the giant radio sources may be a contributing factor, this 
is possibly due to greater dissipation of energy close to the nucleus in 
the sources with low-powered radio jets. This can also be seen in the plot of the total
radio luminosity against the core radio luminosity (Figure 9b). A linear least-squares
fit shows that they are related as logP$_c$ = (0.44$\pm$0.043) logP$_t$ + (12.35$\pm$1.16)
for galaxies and logP$_c$ = (0.59$\pm$0.047) logP$_t$ + (8.58$\pm$1.25) when quasars are 
also included.  The sample
includes core flux density measurements for almost our entire sample of 3CR as well
as giant radio sources.
The trend is consistent with the results reported earlier by Feretti et al. (1984)
and Giovannini et al. (1988) although in their studies many of the cores had reasonably 
high upper limits to the core flux density. A similar trend considering 3CR and B2
sources has been found by de Ruiter et al. (1990).

\begin{figure}
\vbox{
\vspace{-1.0 in}
\psfig{figure=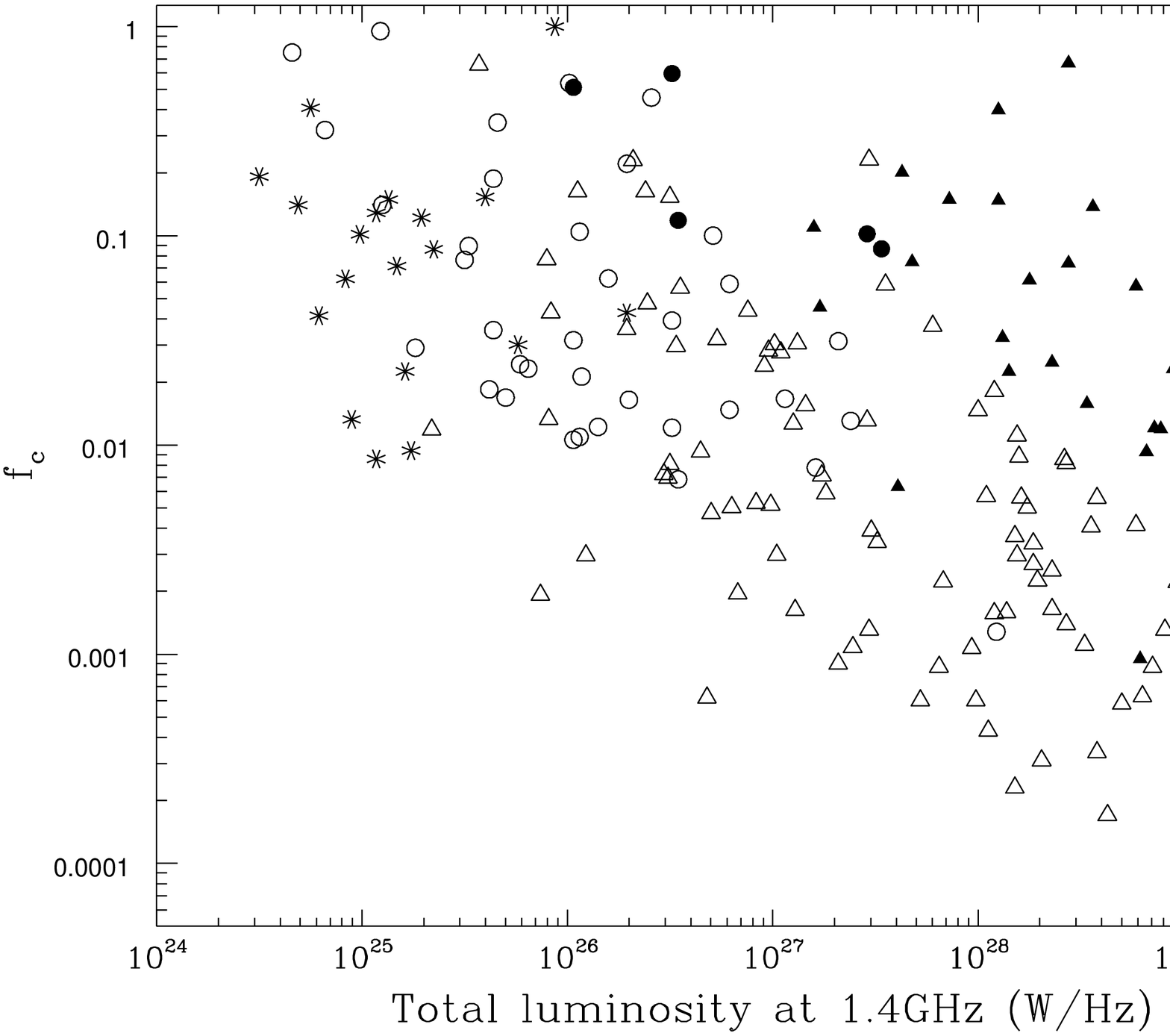,height=3.5in}
\vspace{-0.75 in}
\psfig{figure=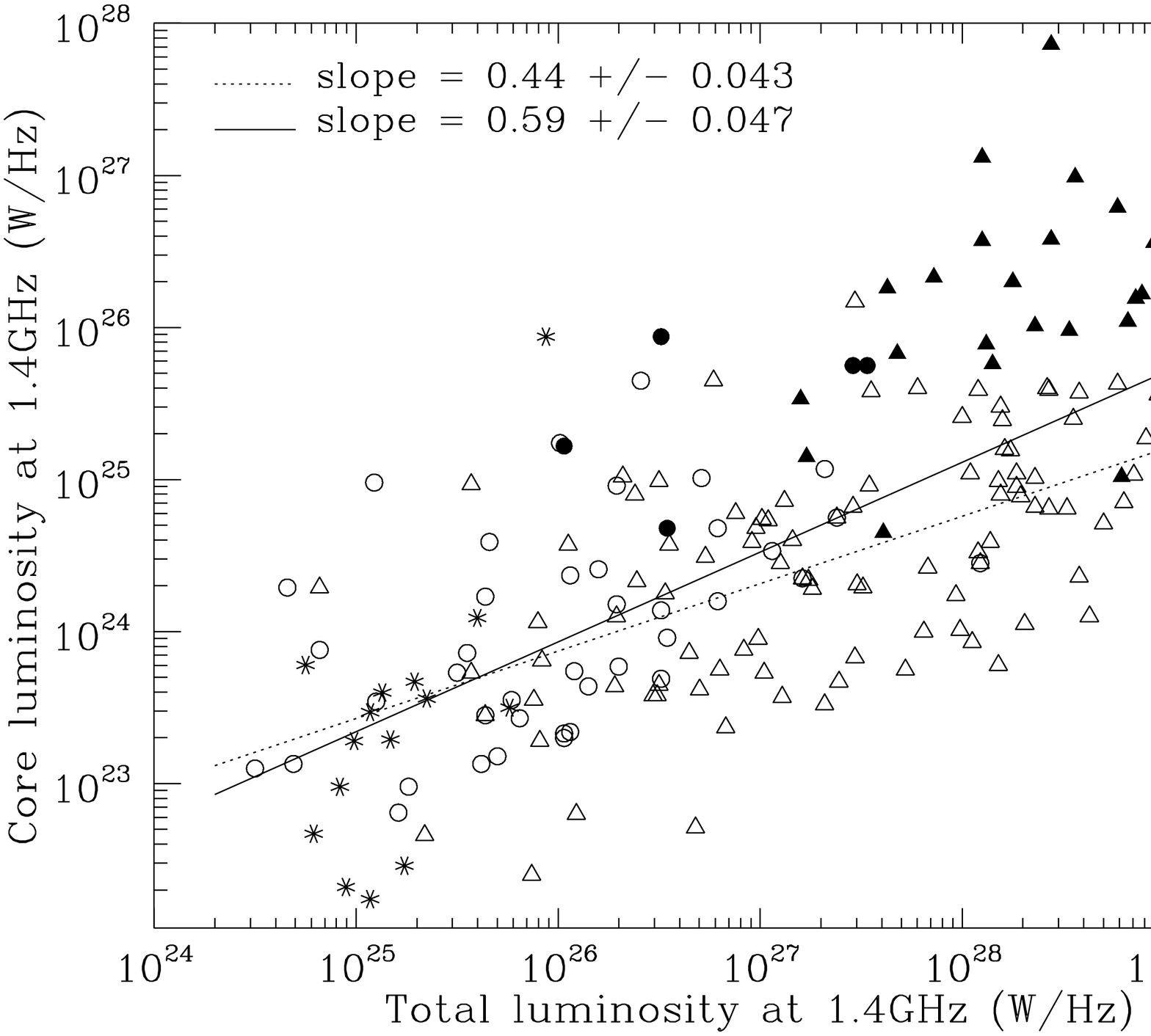,height=3.5in}
\vspace{0.10 in}
}
\caption{{\bf(a)} The fraction of emission from the core and {\bf(b)} 
the core radio luminosity plotted against the
total radio luminosity of the source.
The continuous line logP$_c$ = (0.59$\pm$0.047) logP$_t$
+ (8.58$\pm$1.25) is the linear least-squares fit including both galaxies and
quasars, while the
dotted line represented by logP$_c$ = (0.44$\pm$0.043) logP$_t$ + (12.35$\pm$1.16)
is the linear least-squares fit for galaxies only.
The giant quasars and galaxies
are shown by filled and open circles respectively;
while the 3CR quasars and galaxies are shown by filled and open
triangles respectively, except for the sources with clear FRI structure. The FRI
sources have been marked with an asterisk.
}
\end{figure}

Considering the giant and smaller sources of similar radio luminosity, we do not find
a significant trend for the giant radio sources to have more prominent radio cores
than the smaller ones. For example in the luminosity range, 10$^{25}$ to 10$^{27}$ W Hz$^{-1}$ where
there is maximum overlap of the giants and the smaller sources, the median value
of f$_c$ for the giant galaxies is  0.035 compared to 0.031 for the smaller sources. This also
suggests that the giants are similar objects to the normal radio sources except
for being larger and perhaps older. It is perhaps worth noting that in a couple 
of giant sources such as 3C236 which has a steep-spectrum core, and 1144+352 which has a
GPS core, the high f$_c$ values may be different from the conventional flat-spectrum cores.

\begin{table} \caption{ Core flux densities }
\begin{tabular}{l l l l  }
Source & Epoch   & Flux     & References\\
name   &         & density  &          \\
       &         & (mJy)    &          \\
       &         &          &          \\
NGC315 & 1978.56 & 620$\pm$19 & Bridle et al. 1979  \\
       & 1980.80 & 555$\pm$20 & Rudnick et al. 1986 \\
       & 1980.88 & 590$\pm$30 & Perley 1982 \\
       & 1982.55 & 565$\pm$17 & Saikia et al. in prep \\
       & 1989.29 & 588$\pm$18 & Venturi et al. 1993 \\
       & 1995.83 & 735$\pm$22 & Cotton et al. 1999  \\
       & 1996.36 & 695$\pm$21 & Cotton et al. 1999  \\
       & 1996.77 & 686$\pm$21 & Cotton et al. 1999  \\
       & 1996.84 & 668$\pm$20 & Cotton et al. 1999  \\
       & 1997.53 & 689$\pm$21 & Cotton et al. 1999  \\
       &         &          &          \\
       & 1990.92 & $^\#$588$\pm$18 & Venturi et al. 1993 \\
       & 1994.53 & $^\#$746$\pm$22 & Cotton et al. 1999  \\
       &         &          &          \\
NGC6251& 1980.88 & 650$\pm$30 & Perley 1982 \\
       & 1982.55 & 664$\pm$20 & Saikia et al. in prep \\
       &         &          &          \\
4C74.26& 1986.5  & 420$\pm$10 & Riley et al. 1989  \\
       & 1987.5  & 370$\pm$10 & Riley et al. 1989  \\
       & 1988.0  & 310$\pm$10 & Riley et al. 1989  \\
       & 1988.9  & 328$\pm$ 6 & Pearson et al. 1992  \\
       &         &          &          \\
4C34.47& 1982.27 & 90$\pm$10 & Barthel et al. 1989\\
       & 1983.27 & 90$\pm$5  & Barthel et al. 1989\\
       & 1986.44 & 110$\pm$5  & Barthel et al. 1989\\
       &         &          &          \\
       & 1986.17 & $^*$115$\pm$6  & Hooimeyer et al. 1992 \\
       & 1986.40 & $^*$109$\pm$5  & Hooimeyer et al. 1992 \\
       & 1988.73 & $^*$141$\pm$3  & Hooimeyer et al. 1992 \\
       &         &          &          \\
\end{tabular} \\
$^*$ 10 GHz measurments; $^\#$  8 GHz measurments; other values are at 5 GHz. 
\end{table}

\begin{figure*}
\vbox{
\vspace {-2.05 in}
\hbox{
\hspace{0.2in}
\psfig{figure=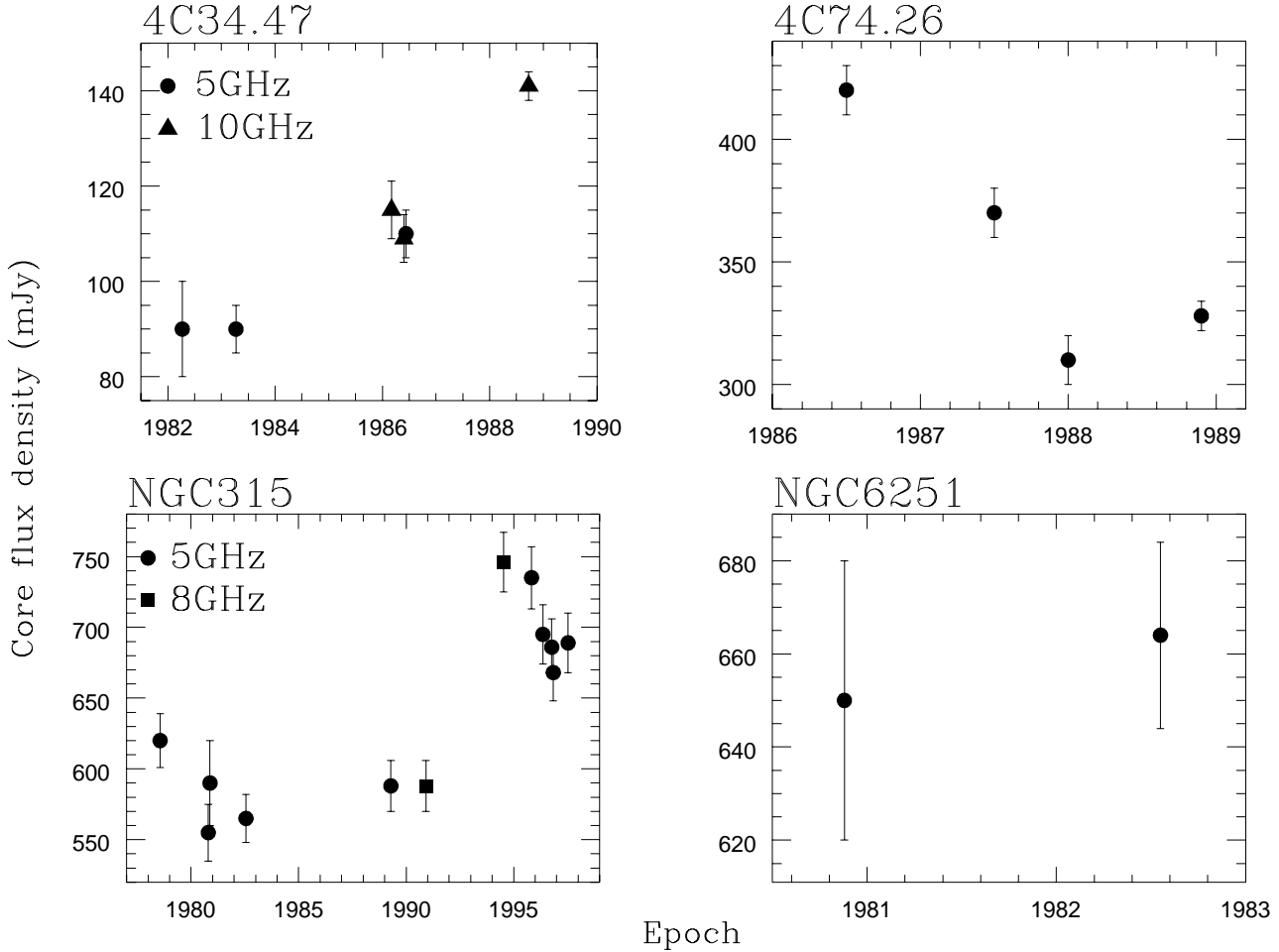,height=7.0in}
}
}
\caption{The flux density of the core at different epochs for the quasars
(a) 4C34.47 and (b) 4C74.26, and the galaxies (c) NGC315 and (d) NGC6251.}
\end{figure*}

\subsection{Core variability}
The clear tendency for the quasar cores to be more prominent than those in galaxies,
even for the giant sources, suggests that the giant quasars are also at small angles to the line
of sight, consistent with the ideas of the unified scheme. The detection of a one-sided
radio jet in the giant quasar 4C74.26 (Riley \& Warner 1990), and superluminal motion in
the quasar 4C34.47 (Barthel et al. 1989; Hooimeyer et al. 1992) also support the unified 
scheme. The inferred 
angle to the line-of-sight for 4C34.47 from the observed superluminal motion is 22$-$44$^\circ$  
(Barthel et al. 1989), implying that the intrinsic size is between 1.5 and 2.7 Mpc. In
the quasar 4C74.26, the inferred orientation angle of 
$\mbox{\raisebox{-0.1ex}{$\scriptscriptstyle \stackrel{<}{\sim}$\,}}$49$^\circ$ (Pearson
et al. 1992) implies an intrinsic size  
$\mbox{\raisebox{-0.1ex}{$\scriptscriptstyle \stackrel{>}{\sim}$\,}}$2 Mpc.

Variability of the core flux density could provide additional test of this scenario
(Blandford \& K\"{o}nigl 1979; Saikia, Singal \& Wiita 1991). 
We have compiled the core flux density of all the giant sources where the epochs of 
measurement are also listed, but find that only four
of them have core flux density measurements at more than one epoch with similar as well as high
angular resolution. Three of these, NGC315, NGC6251 and 4C74.26, have been observed with the
VLA with resolutions of about an arcsec, while for 4C34.47 the core flux density values are 
from VLBI observations with mas resolution. The values for these 4 sources are
listed in Table 4 and plotted against epoch in Figure 10. These are at 5 GHz unless
indicated otherwise in the Table.  For the obsrervations made by Saikia et al. (in preparation),
Bridle et al. (1979), Venturi et al. (1993) and Cotton et al. (1999) we have assumed an error of
3 per cent in the core flux density. Both the quasars 4C74.26 and
4C34.47 exhibit evidence of variability of the core flux density, with 4C34.47 varying by
about 50 per cent over a 4-yr time scale while 4C74.26 has varied by a 
similar amount over a 2-yr time scale. J\"{a}gers et al. (1982) also found the core
of 4C34.47 to be variable from WSRT observations at 5 GHz. Both these quasars have 
prominent cores with 
f$_c$ about 0.60 and 0.51 respectively. Variability information is available for two of the galaxies, 
namely NGC315 and NGC6251, both of which have prominent cores 
with f$_c$ of about 0.32 and 0.75 respectively. NGC315 exhibits no evidence of significant 
variability over a time scale of about 12 yr between 1978 and 1990. The core flux density of 
NGC315 monitored by Ekers, Fanti \& Miley (1983) also showed no evidence of variability although their 
values of core flux density are slightly higher 
due to the poorer resolution of the observations which would have been included a part
of the extended radio jet. However, Cotton et al. (1999) have reported evidence of a flare from 
observations around 1995 at both 5 and 8 GHz. NGC6251 
exhibits no evidence of variability over a time scale of about 2 yr. Although the present 
data are rather limited, this could potentially be an important test and at present provides 
marginal evidence in favour of the unified scheme.

\subsection{Misalignment angle}
A beam of plasma advancing outwards at the same position angle will travel farther 
in a given time scale compared to one whose position angle changes with time. If the 
formation of the giant sources is due to such a steady ejection axis one might find
a statistical difference in the degree of misalignment between the giants and the
smaller sources. However, the misalignment angle, defined to be the supplement of the
angle formed at the core by the outer hotspots, shows no significant difference between
the giants and the smaller sources (Figure 11), the median values being 5 and 6$^\circ$
respectively for the radio galaxies, excluding those with a clear FRI structure. The median
value of $\Delta$ for our Molonglo sample (IC98) is again about 5$^\circ$. The values
are similar when we confine ourselves to objects of similar luminosity in the range of 
10$^{25}$ to 10$^{27}$ W Hz$^{-1}$. The 
3CR quasars exhibit a flatter distribution
which is possibly due to projection effects in sources inclined at smaller angles to the
line-of-sight. The distributions of the misalignment angle also suggest that the giant
sources are basically similar to the smaller ones except for being bigger and 
perhaps older. The hotspot advance speed could depend on more rapid changes in jet 
direction than would be revealed by the overall misalignment angle. This might be 
reflected in the detailed structure of the hotspots including multiple hotspots.
However, to reliably study this aspect one needs high-resolution observations with
a similar number of resolution elements along the source axes for both the giant 
and smaller sources. 

\begin{figure}
\vbox{
\vspace{-1.0 in}
\psfig{figure=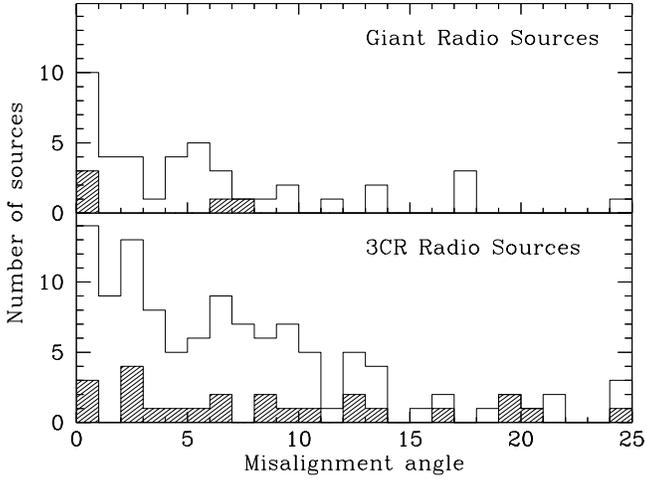,height=3.5in}
}
\caption{The distributions of the misalignment angle, $\Delta$, defined to be the
supplement of the angle formed at the core by the outer hotspots in degrees,
for the samples of giant sources (upper panel) and
the sample of 3CR sources with 50 kpc $<$ D  $<$ 1 Mpc (lower panel). The quasars
are shown hatched in both panels. 
}
\end{figure}

\subsection{Separation ratio}
Although traditionally, the ratio of the separations of the hotspots from the nucleus
for a randomly oriented sample of sources has been used to estimate the hotspot advancement
speed (cf. Longair \& Riley 1979), it has become clear over the years that an asymmetric
environment is also responsible for the observed separation ratios (Saikia 1981; McCarthy,
van Breugel \& Kapahi 1991; Scheuer 1995). There has been some evidence that the compact
steep spectrum sources, which are of subgalactic dimensions have larger separation ratios,
suggesting that they are evolving in an asymmetric environment (Saikia et al. 1995). As the
sources advance outwards into a more symmetric environment on opposite sides and with the 
jets maintaining a constant opening angle, the sources tend to become more symmetric
(Saikia et al. 1996). The sample of 3CR sources considered by Saikia et al. showed some
evidence in support of such a scenario.

Here we attempt to probe the environment on Mpc scales by examining the separation
ratio of the giant radio sources (Figure 12). Considering all the sources in the 
luminosity range of 10$^{25}$ to 10$^{27}$ W Hz$^{-1}$ so that the objects are of
similar luminosity, the median value of the GRGs 
is about 1.39, which is
marginally higher than for the smaller-sized 3CR radio galaxies with projected sizes
between 50 kpc and 1 Mpc, which has a median value
of 1.19. The corresponding values for the entire sample are 1.36 and 1.23 respectively. 
The median value for our Molonglo sample (IC98) is 1.26, which is similar to the 3CR sources.
We have excluded all sources with a clear FRI structure.
Although the difference is only marginally significant, the giant sources do not
appear to be more symmetric and their observed asymmetry could be due to either the beams
encountering density gradients on the scale of Mpcs associated with groups and clusters of 
galaxies after they traverse out of the halo of the parent galaxy, or the jets
traversing outwards with constant jet widths
after they emerge from an asymmetric environment. The jet widths could remain constant if
they are confined by toroidal magnetic fields (Appl 1996, Appl \&  Camenzind 1993a, b). 
One can probe the asymmetry in the environment by examining the relationship between 
jet-sidedness and the separation ratio. There are only 4 galaxies, namely NGC315, 
0319$-$454, 4C74.17 and NGC6251, and 
2 quasars, 4C74.26 and 4C34.47, with well-defined radio jets satisfying the criteria 
suggested by Bridle \& Perley (1984). The jet side is closer in 3 of the 4 galaxies, 
but farther in both the quasars consistent with the environment playing a stronger role
for galaxies while in quasars the effects of orientation seem more significant. The
basic trends expected in the unified scheme are also seen in the giant radio sources. 
The flatter distribution of r$_\theta$ for the 3CR quasars is possibly due
to their smaller angles of inclination to the line-of-sight.

\begin{figure}
\vbox{
\vspace{-1.0 in}
\psfig{figure=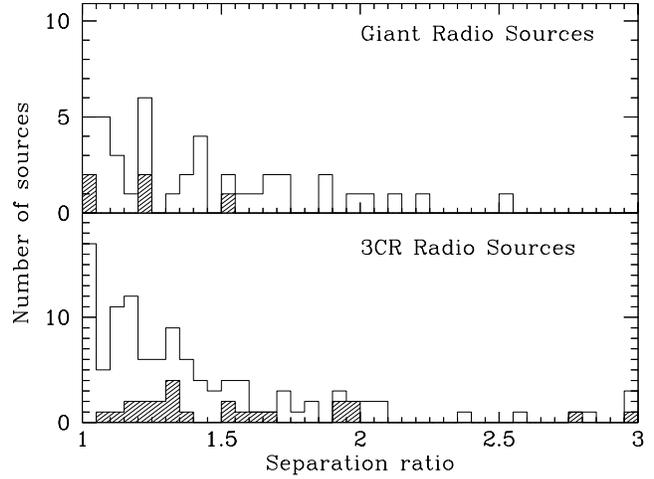,height=3.5in}
}
\caption{The distributions of the separation ratio, r$_\theta$, defined to be the
ratio of the separation of the farther hotspot or peak of emission from the nucleus to that of the 
nearer one on the opposite side, 
for the samples of giant sources (upper panel) and
the sample of 3CR sources with 50 kpc $<$ D  $<$ 1 Mpc (lower panel). The quasars
are shown hatched in both panels. 
}
\end{figure}

\section{Concluding remarks}
We have presented VLA observations of the giant quasars 0437$-$244 and
1025$-$229 from the Molonglo Complete Sample. These sources have well-defined
FRII radio structure, possible one-sided jets, no significant depolarization
between 1365 and 4935 MHz and low rotation measure ($\mid$ RM $\mid <\,$ 20 rad m$^{-2}$).
We have compiled a sample of 53 known giant radio sources from the literature,
and have compared some of their properties with a complete sample of 3CR
radio sources with sizes between 50 kpc and 1 Mpc to investigate the evolution 
of giant sources, and test their consistency with the unified scheme for radio galaxies and
quasars. The conclusions are summarized briefly here. 
\begin{enumerate}
\item The power-linear size or P-D diagram for the 3CR and giant sources show a
deficit of sources with radio luminosity greater than about 2$\times10^{27}$ W Hz$^{-1}$
at 1.4 GHz  and sizes over a Mpc. A similar trend was noted
earlier by Kaiser et al. (1997), and appears to be true for this much larger sample
of giant sources selected from samples covering a wide flux density range.
The location of the giants in the P-D diagram suggests that they have evolved from
the smaller sources. Suggestions that they might be of similar age to the smaller
sources from spectral index studies should be treated with caution because of the
large number of uncertainties and assumptions in these estimates. 
\item The equipartition magnetic field, B$_{eq}$, is smaller than the equivalent 
magnetic field of the microwave background radiation, B$_{ic}$, for the giant sources,
while the reverse is true for the smaller sources. Thus inverse-Compton losses dominate
for the giant radio sources, while synchrotron radiation losses are more important for
the smaller sources. This is likely to severely limit the number of giant radio sources
at large redshifts. 
\item We find an inverse
correlation between the degree of core prominence and total radio luminosity, and
show that the giant radio sources have similar core strengths to the smaller sources
when sources of similar total luminosity are considered. Although many of the giants
have stronger cores than the high-luminosity FRII radio sources (cf. Saikia \& Kulkarni
1994), this is largely due to the inverse correlation between 
the degree of core prominence and total radio luminosity, and does not necessarily
indicate a higher nuclear activity or more powerful central engine. The more prominent
cores in the lower luminosity sources is possibly due  to greater dissipation of energy
by the radio jet close to the nucleus.
\item The degree of collinearity for the giant radio sources is similar to that of the
smaller sources, suggesting that the steadiness of the axis is not the determining factor
for the formation of giant radio sources. 
\item The ratio of separation of the outer hotspots for the giant sources appears marginally
larger than the smaller-sized sources. This is somewhat surprising and could be possibly due
to interaction of the energy-carrying beams with cluster-sized density gradients far from the
parent galaxy. For 6 sources with radio jets, the hotspot on the jet side is closer for 3 of
the 4 galaxies and none of the two quasars. This suggests that the environment plays a stronger role
for galaxies while in quasars the effects of orientation seem more significant.
\item The giant quasars have more prominent cores, one of which 4C34.47 exhibits superluminal
motion and the cores of both quasars with adequate data exhibit evidence of variability. 
Unlike the quasars, the radio cores of one of the galaxies, NGC6251,  exhibits
no evidence of significant variability, while the other NGC315 exhibits evidence of a flare
around 1995 after maintaining a nearly constant flux density for about 20 yr including
the observations of Ekers et al. (1981). Although the
available data are very limited, these are consistent with the unified schemes for 
radio galaxies and quasars.
\end{enumerate}

\section*{Acknowledgments}
We thank Sivakumar Manickam and Divya Oberoi for computational help, an anonymous referee
for meticulously reading the manuscript and making several helpful suggestions and comments, Jayaram
Chengalur and Kandaswamy Subramanian for their comments on the  manuscript
and many of our colleagues for useful discussions.
The National Radio Astronomy Observatory  is a
facility of the National Science Foundation operated under co-operative
agreement by Associated Universities Inc. We thank
the staff of the Very Large Array for the observations. 
This research has made use of the NASA/IPAC extragalactic database (NED)
which is operated by the Jet Propulsion Laboratory, Caltech, under contract
with the National Aeronautics and Space Administration.

{}

\end{document}